\documentclass[12pt,reqno]{amsart}
\usepackage{amssymb,epsfig,amsfonts,amsmath,amsthm}
\usepackage{verbatim}

\newcommand{\bC}{{\mathbb C}}
\newcommand{\bR}{{\mathbb R}}

\newcommand{\cD}{\mathcal{D}}

\newcommand{\vV}{\mathbf{V}}
\newcommand{\vW}{\mathbf{W}}
\newcommand{\vU}{\mathbf{U}}
\newcommand{\vT}{\mathbf{T}}
\newcommand{\vD}{\mathbf{D}}
\newcommand{\ve}{\mathbf{e}}
\def\phase{\boldsymbol{\varphi}}
\newcommand{\vf}{\mathbf{f}}
\newcommand{\vk}{\mathbf{k}}
\newcommand{\vn}{\mathbf{n}}
\newcommand{\vm}{\mathbf{m}}
\newcommand{\vr}{\mathbf{r}}

\newcommand{\vw}{\mathbf{w}}

\newcommand{\vomega}{\boldsymbol{\omega}}

\newcommand{\vpsi}{\boldsymbol{\psi}}

\newcommand{\A}{\boldsymbol{\mathcal A}}
\newcommand{\K}{\boldsymbol{\mathcal K}}
\newcommand{\vgamma}{{\boldsymbol{\gamma}}}

\newcommand{\M}{{\mathcal M}}
\newcommand{\RiemB}{\mathcal B}
\newcommand{\interchange}{\iota}
\newcommand{\calR}{\mathcal R}
\newcommand{\calT}{\mathcal T}

\newcommand{\ZZ}{{\mathbb Z}}
\newcommand{\R}{{\mathbb R}}
\newcommand{\ulA}{\underline{A}}
\newcommand{\ula}{\underline{a}}
\newcommand{\ulb}{\underline{b}}
\newcommand{\ull}{\underline{\ell}}
\def\transpose#1{{}^t\negthinspace{#1}}
\def\widebar{\overline}
\def\blamda{\widebar{\lambda}}
\newcommand{\Mt}{\transpose{M}}
\def\realpart{\operatorname{\sf Re}}

\newcommand{\tr}{\operatorname{tr}}

\newcommand{\sn}{\operatorname{sn}}
\newcommand{\dc}{\operatorname{dc}}
\newcommand{\dn}{\operatorname{dn}}
\newcommand{\cn}{\operatorname{cn}}
\newcommand{\cs}{\operatorname{cs}}

\newlength{\tricklength}

\newcommand{\bel}[2]{\begin{equation}\label{#1} #2 \end{equation}}

\newcommand{\rd}{\mathrm{d}}            
\newcommand{\re}{\mathrm{e}}            
\newcommand{\ri}{\mathrm{i}}            

\newtheorem{theorem}{Theorem}
\numberwithin{theorem}{section}

\newtheorem{cor}[theorem]{Corollary}
\newtheorem{lemma}[theorem]{Lemma}
\newtheorem{lem}[theorem]{Lemma}
\newtheorem{proposition}[theorem]{Proposition}
\newtheorem{prop}[theorem]{Proposition}

\theoremstyle{definition}
\newtheorem{definition}[theorem]{Definition}

\newtheorem{remark}[theorem]{Remark}

\begin{document}
\title{Finite-gap Solutions of the Vortex Filament Equation, I}
\author{A. Calini and T. Ivey} 
\thanks{The authors were partially funded by NSF grant DMS-0204557}
\address{Department of Mathematics, College of Charleston \\ Charleston SC 29424 USA}
\email{calinia@cofc.edu, iveyt@cofc.edu}

\date{\today}

\maketitle
\begin{abstract}
For the class of quasi-periodic solutions of the vortex filament equation we study  connections between the algebro-geometric data used for their explicit construction,  and the geometry of the evolving curves. We give a complete description of genus one solutions,  including  geometrically interesting special cases such as Euler elastica, constant torsion curves, and self-intersecting filaments. We also prove generalizations of these connections to higher genus. \end{abstract}
\section{Introduction}
\newcommand{\fsu}{\mathfrak{su}}

In 1972, H. Hasimoto~\cite{Ha} introduced the complex function
\bel{Hmap}{
q = \tfrac12\, \kappa \exp\left({\ri \int \tau\, \rd s}\right)
}
of the curvature $\kappa$ and the
torsion $\tau$ of a space curve,
and showed that if the position vector $\vgamma$ of the curve evolves according to the vortex filament equation (VFE),
\begin{equation}
\label{VFE}
\frac{\partial \vgamma}{\partial t}= \frac{\partial \vgamma}{\partial s}\times \frac{\partial^2 \vgamma}{\partial s^2},
\end{equation}
then the {\em potential} $q(s,t)$ is  a solution of the
focussing cubic nonlinear  Schr\"{o}dinger equation (NLS)
\begin{equation}
\label{NS2}
\ri q_t + q_{ss} + 2 |q|^2 q = 0.
\end{equation}
In these equations, $s$ is the arclength parameter of the curve and the spatial variable for the NLS potential,
and $t$ is  time.
In fact, the evolving curve defines
$q(s,t)$ uniquely, up to multiplication by a unit modulus constant.

The association $\vgamma(s,t) \mapsto q(s,t)$ is often referred
to as the {\em Hasimoto map}.
Since the NLS had been shown to be completely integrable
by Zakharov and Shabat \cite{ZS,FT}, Hasimoto's discovery meant that the
VFE is also completely integrable, and possessed all the attendant
structure: infinite sequences of commuting flows and common conservation laws,
solution by inverse scattering, and special solutions such as solitons and,
more generally, finite-gap solutions.

The VFE was proposed a century ago as a simplified model
for the self-induced motion of vortex lines in an incompressible fluid \cite{Ri,Lam}.
Hasimoto's discovery has developed into one of the richest
examples of the connection between curve geometry and integrability.
The understanding of this connection has progressed in recent years along different directions. In the case of the VFE, its bihamiltonian structure and recursion operator have been given a geometrical realization \cite{LP, Sa}  and  its relation to the NLS has been shown to possess a natural geometric interpretation \cite{SY14, Ca}.  The study of special solutions, including solitons, finite-gap solutions and their homoclinic orbits has been undertaken by several researchers, employing techniques ranging from perturbation theory \cite{Ke, CI3}  to methods of algebraic-geometry and B\"acklund transformations \cite{Sy,CGS, CI1}, in particular with the aim of understanding whether the presence of infinitely many symmetries and integral invariants implies restrictions on the topology of the evolving curves \cite{MR, Br}.

In this paper, we focus on connections between geometric information about the evolving
curve $\vgamma$, and features of the NLS potential $q(s,t)$.  Specifically, we limit ourselves
to the class VFE solutions that correspond to periodic and quasiperiodic
finite-gap solutions of NLS, which (as explained
in \S\ref{thetasection}) may be constructed using algebro-geometric data, and it is this
data which we seek to connect with the geometry (and topology) of $\vgamma$.
Note that this choice is particularly significant for closed curves, since the finite-gap solutions are dense in the space of all periodic solutions \cite{Gri}, and it is therefore important to study their geometrical and topological properties.

Of course, the class of curves that correspond to finite-gap solutions of the NLS only makes sense if the Hasimoto map is invertible.  In fact, the inverse map is defined using the solutions of the AKNS system for \eqref{NS2}.
As is well-known, the AKNS system consists of a pair of first-order linear systems
\begin{equation}
\label{LINS}
\begin{aligned}
\vpsi_s &= U \vpsi \\
\vpsi_t & = V \vpsi,
\end{aligned}
\qquad \vpsi(s,t) \in \bC^2
\end{equation}
for a vector-valued function $\vpsi$, for which the solvability
or ``zero curvature condition'' $\vpsi_{st}=\vpsi_{ts}$ is equivalent to \eqref{NS2}.
Expressed in terms of the Pauli matrix
$\sigma_3 = \begin{pmatrix}1 & 0 \\ 0 & -1 \end{pmatrix}$, the matrices on the
right-hand sides in \eqref{LINS} are
$$
U =-\ri\lambda \sigma_3 +  \begin{pmatrix}0 & \ri q \\ \ri \bar{q} & 0 \end{pmatrix},
\qquad
 V = \ri (|q|^2-2\lambda^2) \sigma_3 +
\begin{pmatrix} 0 & 2\ri \lambda q - q_s \\
 2\ri \lambda \bar{q} +\bar{q}_s & 0 \end{pmatrix}.
$$
These matrices depend on $s$ and $t$ through the complex-valued
function $q$, and on the {\em spectral parameter} $\lambda$.
(In equations \eqref{Hmap},\eqref{NS2} and \eqref{LINS}, and
elsewhere in this paper, our conventions
have been chosen so as to be consistent with
the important monograph by Belokolos, Bobenko {\it et al.} \cite{BBEIM}.)
Because the first equation of \eqref{LINS} can be written as $L\vpsi = \lambda \vpsi$
for a first-order matrix differential operator $L$, $\vpsi$ is known as
an {\em eigenfunction} associated with $q$.

If one can compute a common solution of \eqref{LINS} for a given potential $q(s,t)$,
then the curve that corresponds to $q$ under the Hasimoto map is obtained
by means of the following reconstruction procedure:
\begin{proposition}[after A. Sym \cite{SY14}, K. Pohlmeyer \cite{POL}]
Let $q(s,t)$ satisfy \eqref{NS2},
let $\Psi(s,t;\lambda)$ be a matrix whose columns are linearly independent solutions of \eqref{LINS},
such that $\Psi(0,0;\lambda)$ is a fixed matrix in $SU(2)$, and
let
\begin{equation}
\label{RECO}
\Gamma(s,t)=\left. \Psi^{-1}\frac{\rd \Psi}{\rd\lambda} \right|_{\lambda=0}.
\end{equation}
Then $\Gamma$ satisfies \eqref{VFE}, and the curvature and torsion of $\Gamma$ satisfy \eqref{Hmap},
assuming we identify $\fsu(2)$ with $\R^3$ via a
linear isomorphism $F$, specified by
$$F:\begin{pmatrix}-\ri & 0 \\ 0 &\ri\end{pmatrix}\mapsto \ve_1, \qquad
F: \begin{pmatrix}0 & 1\\ -1 & 0 \end{pmatrix}\mapsto \ve_2, \qquad
F: \begin{pmatrix}0 & \ri \\ \ri & 0 \end{pmatrix}\mapsto \ve_3.
$$
\end{proposition}
\noindent
(Of course, the reason this construction makes sense is that
$U,V$ take value in $\fsu(2)$ when $\lambda \in \R$.)
Thus, if we begin with a solution of \eqref{VFE}, construct the potential $q(s,t)$, and solve the
linear system, then the curve given by \eqref{RECO} will differ from what we started with
by an isometry of $\R^3$.

Formula \eqref{RECO} has become known as the Sym-Pohlmeyer reconstruction
formula.  More generally, if in \eqref{RECO} we evaluate at an arbitrary real value of
$\lambda$, i.e.,
\begin{equation}
\label{generalSymformula}
\Gamma=\left. \Psi^{-1}\frac{\rd \Psi}{\rd\lambda} \right|_{\lambda=\Lambda_0}, \qquad \Lambda_0 \in \R,
\end{equation}
then $\Gamma$ still satisfies \eqref{VFE}, but with a potential that differs from what we
started with by the transformation $q(s,t) \mapsto \re^{\ri(a s-a^2 t)} q(s-2a t,t)$, $a\in \R$, which
preserves solutions of \eqref{NS2}.  We will have recourse to the generalized Sym-Pohlmeyer formula
\eqref{generalSymformula} in order to associate a closed curve to a periodic NLS potential.

By differentiating \eqref{generalSymformula} with respect to $s$, we see that the unit tangent vector
$\vT$ along $\vgamma$ and a pair of unit normal vectors $U_1,U_2$ are given by
$$\vT = \Psi^{-1}\begin{pmatrix}-\ri & 0 \\ 0 &\ri\end{pmatrix}\Psi,
\qquad
\vU_1 = \Psi^{-1}\begin{pmatrix}0 & 1\\ -1 & 0 \end{pmatrix}\Psi,
\qquad
\vU_2 = \Psi^{-1}\begin{pmatrix}0 & \ri \\ \ri & 0 \end{pmatrix},
$$
with the aforementioned identification of $\fsu(2)$ and $\R^3$ understood.  These
vectors do not constitute a Frenet frame for $\vgamma$, but rather satisfy
the {\em generalized natural frame equations}:
$$\dfrac{\rd}{\rd s} \begin{bmatrix} \vT \\ \vU_1 \\ \vU_2 \end{bmatrix}
= \begin{bmatrix} 0 & k_1 & k_2 \\ -k_1 & 0 & \Lambda_0 \\ -k_2 & -\Lambda_0 & 0 \end{bmatrix}
\begin{bmatrix} \vT \\ \vU_1 \\ \vU_2 \end{bmatrix},
$$
where $k_1,k_2$ are the {\em natural curvatures}, which in this case coincide with the
real and imaginary parts of $2q$, respectively.  (If $\Lambda_0=0$, then $(\vT,\vU_1,\vU_2)$
comprise a {\em natural} or {\em relatively parallel adapted frame} \cite{Bishop} along $\vgamma$.)
The interpretation of the Hasimoto map in terms of natural curvatures, and the connection
between the eigenfunction matrix $\Psi$ and the natural frame, will be used when we discuss
closure conditions in \S\ref{closure}.

\bigskip
\subsubsection*{Organization of the paper}
In \S\ref{thetasection} we introduce the notion of finite-gap solutions,
and give formulas for the eigenfunctions, potential and curve in terms of
the algebro-geometric data---specifically a hyperelliptic Riemann
surface $\Sigma$ and a positive divisor $\cD$ on it, satisfying
certain reality and genericity conditions.
We also use these formulas to give conditions under
which the curve is smoothly closed, and conditions under which it
periodically intersects itself.  (The closure conditions duplicate
results of Grinevich and Schmidt \cite{GSc}, but here we derive them
directly from the curve formulas.)

In \S\ref{genusone} the ideas of the
previous section are worked out in detail for the genus one case.
In particular, we show that  the resulting
curves  are centerlines for Kirchhoff elastic rods.
We study the geometric features of curves associated with
special configurations of the algebro-geometric data:
we show that
the elastic rod centerline 
becomes an Eulerian  elastic curve precisely when the zeros of a certain
meromorphic differential involved in the finite-gap construction
coincide, and that the centerlines have constant torsion exactly
when the branch points of the
elliptic curve have an extra axis of symmetry.  (They are always
symmetric under complex conjugation.)

In \S\ref{symmetric}, we discuss some generalizations of
this phenomenon---symmetric branch points giving
rise to special geometric features of the curve---in higher genus.
In an appendix \S\ref{appendix} we relate our eigenfunction formulas to
those given in \cite{BBEIM}, and give a straightforward
derivation of the reality conditions, as well as some new
formulas for calculating the algebro-geometric data.

\subsubsection*{Forthcoming papers in this series}
One of our main interests is the extent to which topological
information (in particular, knot type) can be ``read off'' from
the algebro-geometric data that gives rise to a closed filament.
However, it is difficult to obtain examples of closed finite-gap
filaments to base conjectures on in the first place.
As will become apparent in \S\ref{closure}, the conditions
under which the filament is smoothly closed are difficult to
compute in general.

One alternative to tackling the closure conditions head-on is
to begin with a closed curve and deform its spectrum (in particular,
the branch points of the hyperelliptic curve) in a way
that preserves closure.  Deformations that preserve the periodicity
of $q$ have been studied by Grinevich and Schmidt \cite{GSa}.  In our
next paper, we will examine a special adaptation of these deformations
that also preserves closure.  We will show that, by starting with
a multiply-covered circle (corresponding to a simple plane wave
solution of NLS) and applying successive deformations, we obtain a
{\em labelling scheme} that matches the deformation data with
the knot type of the resulting filament, assuming the sizes of
the deformations are sufficiently small.

\section{Finite-gap Solutions}
\label{thetasection}

Soliton equations, when considered on periodic domains, admit  classes of
special solutions which are the analogues of solitary waves for rapidly
decreasing initial data on an infinite domain.
An N-phase, or finite-gap, quasiperiodic potential is an NLS solution of the form
$$q(s,t)=q(\theta_1, \dots , \theta_N), \qquad \theta_i(s,t)=k_is+w_i t, \ i=1, \ldots , N,$$
such that $q$ is periodic   in each phase: $q(\theta_1, \ldots, \theta_i+2\pi, \dots , \theta_N) = q(\theta_1, \ldots, \theta_i, \ldots , \theta_N)$. The vector $\vk=(k_1, \dots , k_N)$ is called the vector of spatial frequencies, and $\vw=(w_1, \dots , w_N)$ the vector of time frequencies.

Finite-gap solutions of soliton equations such as KdV, NLS, and general AKNS
systems were first constructed 
in the mid-1970's (see the Introduction in \cite{BBEIM} for references).
In particular, finite-gap solutions to the nonlinear Schr\"odinger equation
were first constructed by Its \cite{Its} and Its and Kotlyarov \cite{IK}.
Here, we use Krichever's construction~\cite{Kr} based on the Baker-Akhiezer eigenfunction, and the Sym-Pohlmeyer reconstruction formula~(\ref{RECO}) to derive the explicit formulas for finite-gap solutions of the VFE (\ref{VFE}). We mostly follow the notation and the discussion of reality conditions in~\cite{BBEIM}.

A finite-gap  solution is defined by a set of data on  a smooth hyperelliptic Riemann surface $\Sigma$ of genus $g$, with distinct branch points $\{\lambda_i\}_{i=1}^{2g+2}$, described by the equation
\begin{equation}
\mu^2 = \prod_{i=1}^{2g+2} (\lambda - \lambda_i).
\end{equation}
Let $\pi  : \Sigma \rightarrow \bC \cup \{ \infty \}$ be the standard
hyperelliptic projection defined by $\pi(P)=\lambda$, with $P=(\lambda, \mu) \in \Sigma$, and let
$\infty_+ +\infty_- $ be the pole divisor of $\pi$
(labelled so   that $\mu^{-1}\lambda^{g+1} \rightarrow \pm 1$ as $\lambda \rightarrow  \infty_\pm$).
Let ${\cD} = P_1 + \cdots + P_{g+1}$ be a non-special divisor  of degree $g+1$
and not containing  $\infty_{\pm}$.
Recall that a positive  divisor $\cD$ of degree at least $g$
is called non-special when the linear space of meromorphic functions with poles in
$\cD$ has dimension $l(\cD)=\deg(\cD)-g+1$. The special divisors,  which are those such that $l(\cD)>\deg(\cD)-g+1,$
are the critical points of the Abel map.

Krichever's main  idea is to construct a function $\psi (P)$ on
$\Sigma$ which is uniquely defined by a given set of singularities and by  prescribed
asymptotic behavior near
$\infty_\pm$.

\begin{definition}
A  Baker-Akhiezer function associated to $(\Sigma , \cD,\infty_{\pm})$ is defined by the following
properties:
\begin{itemize}
\item
$\psi$ is meromorphic on $\Sigma \backslash \infty_{\pm}$ and has
pole divisor contained in $\cD$,
\item
$\psi$  has essential singularities at $\infty_{\pm}$ that locally are of the form

$$\psi  \sim e^{\pm \ri (\lambda s + 2 \lambda^2 t)} \left[ c(s,t)+O(\lambda^{-1})\right]$$

 for $P\rightarrow \infty_\pm$ and  $\lambda=\pi (P)$, where $c(s,t)$ is a function of  arbitrary  complex parameters $s$ and $t$.
\end{itemize}
\end{definition}

The singular structure of $\psi$ and its normalization at the essential singularities define it uniquely as described in the following
\begin{proposition}[Krichever~\cite{Kr}]
Suppose that the following technical condition holds:
\vspace{.3cm}

{\bf \slshape Genericity Condition:}
{\rm
The non-special divisor $P_1 + \cdot \ \cdot \ \cdot \ + P_{g+1} - \infty_{+} -
\infty_{-}$ is not linearly equivalent to a positive divisor.
}
\vspace{.3cm}

\noindent
Then, for $|s|,|t|$ sufficiently small, the linear vector space of  Baker-Akhiezer eigenfunctions
associated with $(\Sigma , \cD,\infty_{\pm})$ is $2$-dimensional and has
a unique basis  $\psi^1$ and $\psi^2$, such that the vector-valued function
 $\displaystyle \vpsi=\left( \begin{array}{c} \psi^1 \\ \psi^2 \end{array} \right)$ has  the following normalized expansions at $\infty_-$ and $\infty_+$, respectively:
 \begin{equation}
\vpsi=e^{-\ri(\lambda s+2\lambda^2t)}\left[ \left( \begin{array}{c} 1 \\ 0 \end{array} \right) +O(\lambda^{-1})\right], \qquad \vpsi=e^{\ri(\lambda s+2\lambda^2t)}\left[ \left( \begin{array}{c} 0 \\ 1 \end{array} \right) +O(\lambda^{-1})\right] .
\end{equation}
\end{proposition}

\noindent
{\bf Remarks:}

1. The Genericity Condition can be rephrased as, there exists no non-constant
meromorphic function with pole divisor $P_1 + \ ...\ + P_{g+1}$ which vanishes
simultaneously at $\infty_+$ and at $\infty_-$.

2. The proof of uniqueness follows from the fact that the ratio $\psi^1/\psi^2$ of two Baker-Akhiezer functions is a meromorphic function whose poles lie in the zero divisor of $\psi^2$. The condition of non-speciality ensures that the dimension of the space of such functions is 2. Since $l(\cD-\infty_+)=l(\cD-\infty_-)=1$, we can normalize $\psi^1$ and $\psi^2$ by requiring that they vanish at $\infty_+$ and $\infty_-$, respectively. Because $l(\cD-\infty_+ - \infty_-)=0$, which follows from the Genericity Condition, they must be independent. For more details, see  \S\ref{realityconds}.

\begin{figure}[ht]
\centering
\includegraphics[width=4.2in]{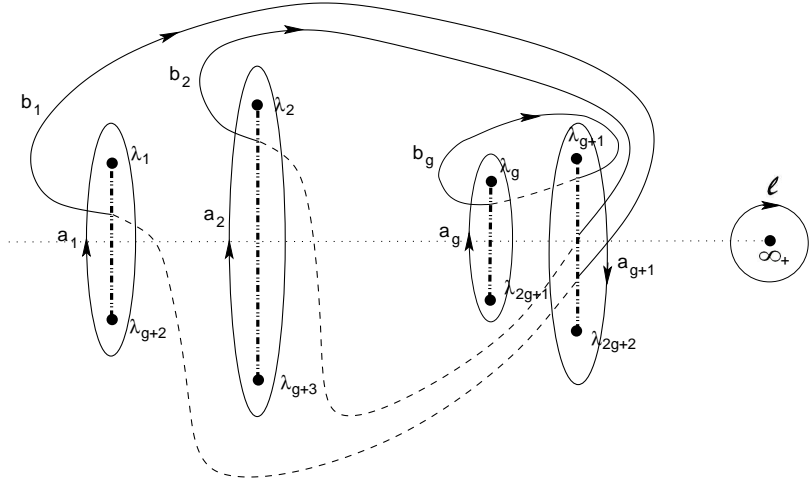}
\caption{Homology cycles on $\Sigma$, adapted from \cite{BBEIM}.
The dashed vertical lines are branch cuts, the solid curves are
portions of homology cycles on the upper sheet, the
dashed curves are portions on the lower sheet, and the
dotted horizontal line is the real axis in the $\lambda$-plane.  Also shown
are cycles $a_{g+1}$ and $\ell$, used in \S\S\ref{symmetric} and
\ref{appendix}.
}\label{genbasis}
\end{figure}

\bigskip
The existence of the Baker-Akhiezer eigenfunction is shown by explicitly constructing its components in
terms of the Riemann theta function of $\Sigma$. We summarize the main steps of this algebro-geometric construction. (See \cite{BBEIM,Du} for a comprehensive discussion and application of this method to several other soliton equations.)
Let $a_1, \ ... \ , a_g, \, \ b_1, \ ... \ , b_g,$
be a  homology basis for  $\Sigma$, such that
$a_i \cdot a_j=0, \, b_i \cdot b_j =0,\, a_i \cdot b_j = \delta_{ij}$ (see Figure \ref{genbasis}).

\noindent
Let $\omega_1, \ ... \ \omega_g$ be holomorphic differentials on $\Sigma$, normalized as follows
$$
\oint_{a_k}\omega_j =2\pi \ri \delta_{jk}, \qquad j,k = 1,\ ... ,\ g.
$$
We introduce the period matrix $\RiemB$ of entries
\begin{equation}
\label{pmatrix}
\RiemB_{jk}= \oint_{b_k}\omega_j , \qquad j,k = 1,\ ... \ , g,
\end{equation}
and construct the associated Riemann theta function
\begin{equation}
\label{RITHE}
\theta (\mathbf{z}) = \sum_{\mathbf{n} \in {\ZZ}^g} \exp  \left( \tfrac{1}{2}
 \langle \mathbf{n},\RiemB\mathbf{n}\rangle  +  \langle \mathbf{n},\mathbf{z}\rangle
\right) , \quad  \mathbf{z}\in \bC^g .
\end{equation}
The series (\ref{RITHE}) is absolutely convergent; this follows from the fact that $\realpart(\RiemB)$ is a negative definite matrix. Moreover, if  $\ve_k$'s  are the standard basis vectors of $\bC^g$ and
$\vf_k=\RiemB\ve_k,$ for $k=1, \ldots g$, then
$$
\theta(\mathbf{z}+ 2\pi \ri \ve_k)=\theta(\mathbf{z}), \qquad \quad  \theta(\mathbf{z}+\vf_k)=\re^{-\frac{1}{2}\RiemB_{kk}- z_k} \theta(\mathbf{z});
$$
in other words, $\theta$ is a quasi-periodic function. (The $\ve_k$'s are called {\sl periods} and the $\vf_k$'s the {\sl quasiperiods} of $\theta$.) 

We also introduce the Abel map, which takes a divisor  $\sum_{k=1}^g Q_k$ to a point of the complex torus $\operatorname{Jac}(\Sigma)=\bC^g / \Lambda$, where  $\Lambda$ is the $2g$-dimensional lattice spanned by the columns of
the matrix $({\textup I}\, | \RiemB)$.
The Abel map of $\Sigma$ is defined by
\begin{equation}
\A \,: \, \sum_{k=1}^g Q_k   \longrightarrow \sum_{k=1}^g \int_{P_0}^{Q_k}
\vomega \ \mathrm{mod} \Lambda,
\end{equation}
where $P_0 \in \Sigma$ is a given base point.

\subsection{Eigenfunction formulas}
\label{eigenfunctionsection}

\begin{figure}[ht]
\includegraphics[width=4in]{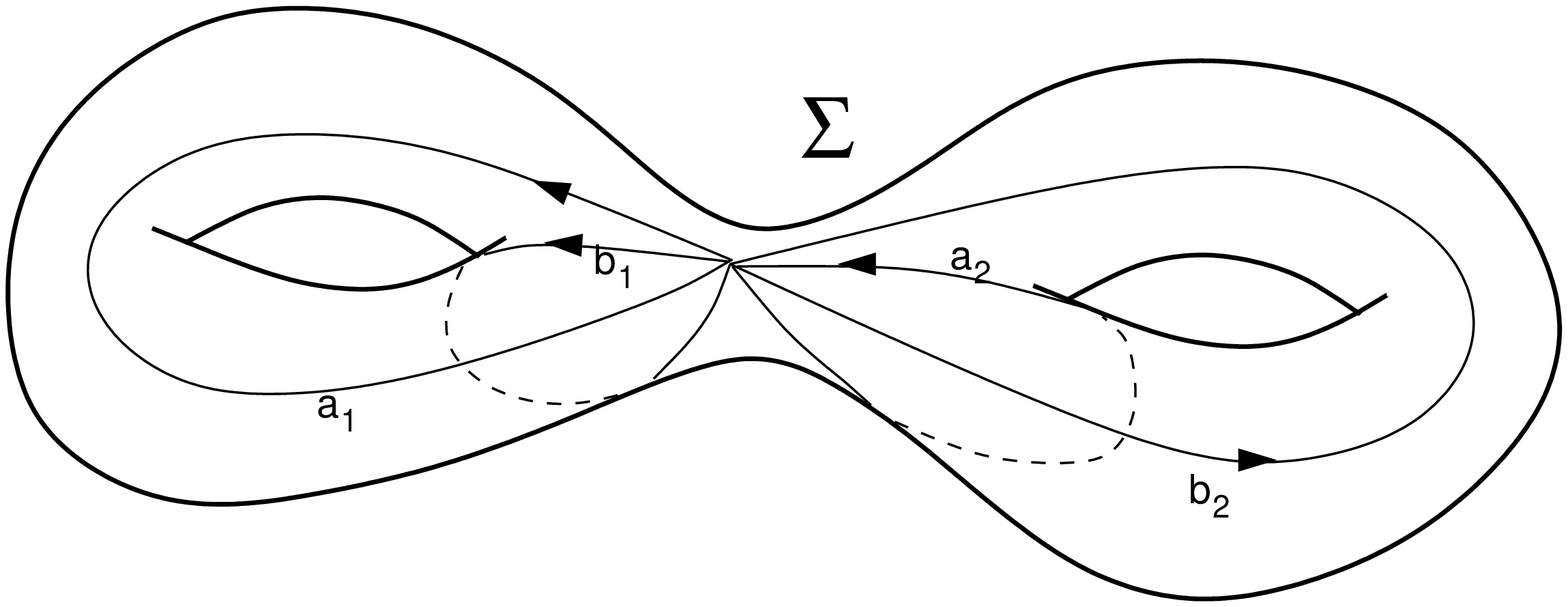}\hspace{.3in}
\includegraphics[height=2in]{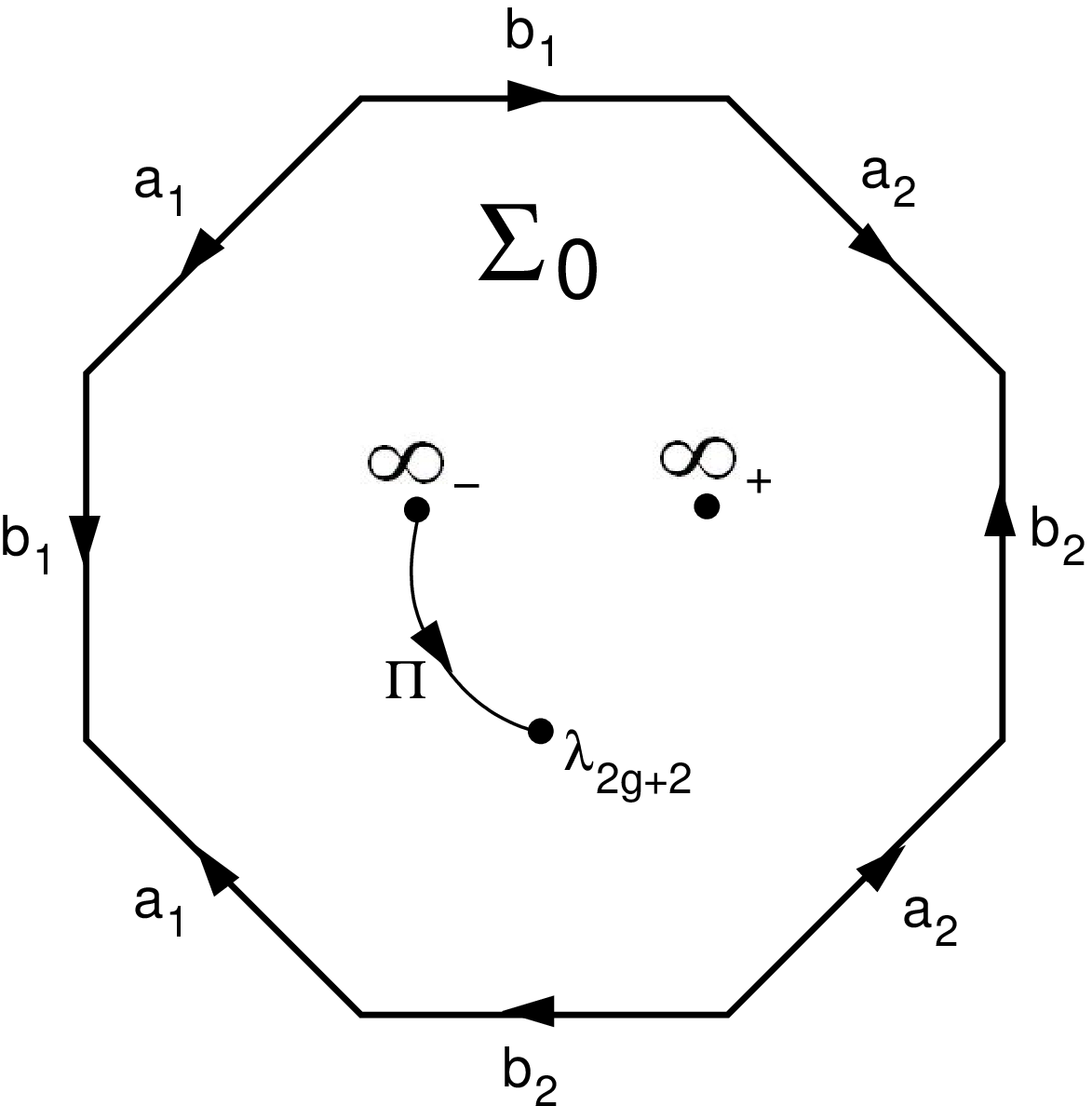}
\caption{Cutting a two-hole torus $\Sigma$ along a homology basis, to yield
the cut surface $\Sigma_0$}\label{cutting Sigma}
\end{figure}

The components of the vector-valued Baker-Akhiezer eigenfunction $\vpsi$ are constructed in terms of ratios of Riemann theta functions:
\begin{equation}
\label{THETA1}\begin{split}
\psi^{1} (P; s, t) = & \exp \left( \ri s ( \Omega_1(P) - \tfrac{E}{2})
+  \ri t ( \Omega_2(P) +\tfrac{N}{2} ) \right)
 \frac{ \theta(\A(P) + \ri \vV s + \ri \vW t  - \A({\cD}_+ )-\K)}
{ \theta(\A(P)  -\A({\cD}_+ )-\K)} \\
& \times \frac{ \theta(\A(\infty_-)  - \A({\cD}_+ )-\K)}
{ \theta(\A(\infty_-) + \ri \vV s + \ri \vW t  - \A ({\cD}_+ )-\K)}
\times \frac{g_+ (P)}{g_+ (\infty_- )},
\end{split}
\end{equation}
and
\begin{equation}
\label{THETA2}\begin{split}
\psi^{2} (P; s, t) = & \exp \left( \ri s ( \Omega_1(P) + \tfrac{E}{2})
+  \ri t ( \Omega_2(P) - \tfrac{N}{2} ) \right)
 \frac{ \theta(\A(P) + \ri \vV s + \ri \vW t  - \A({\cD}_- )-\K)}
{ \theta(\A(P)  -\A({\cD}_- )-\K)} \\
& \times \frac{ \theta(\A(\infty_+)  - \A({\cD}_- )-\K)}
{ \theta(\A(\infty_+) + \ri \vV s + \ri \vW t  - \A ({\cD}_- )-\K)}
\times \frac{g_- (P)}{g_- (\infty_+ )}.
\end{split}
\end{equation}
The ingredients of this formulas are described below:
\begin{enumerate}
\item The essential singularities of $\vpsi$ at $\infty_\pm$ are introduced by means of the unique Abelian differentials
$\rd\Omega_1(P)$ and $\rd\Omega_2(P),$  defined by their asymptotic expansions at
$\infty_\pm$
$$\rd\Omega_1 \sim \pm \rd\lambda, \quad \rd\Omega_2 = \pm 4 \lambda \, \rd\lambda, \quad \text{as} \quad P \rightarrow \infty_\pm,$$ and normalized so that their integrals vanish along the $a$-cycles.

\item We will choose $P_0=\infty_-$ as basepoint for the Abel map $\A$, and
choose the rightmost branch point $\lambda_{2g+2}$ in the lower half-plane as basepoint for
the Abelian integrals $\displaystyle \Omega_j(P)=\int_{\lambda_{2g+2}}^P \rd \Omega_j$.
This latter choice gives the  property
$$\Omega_j \circ \interchange = -\Omega_j,$$
where $\interchange:(\lambda,\mu)\mapsto (\lambda,-\mu)$ is
the sheet interchange automorphism. In this way, the asymptotic expansions for $\Omega_1$ and $\Omega_2$ at $\infty_\pm$ are
$$\displaystyle \Omega_1=\pm(\lambda -\tfrac{E}{2} +o(1))\quad  {\rm and} \quad \Omega_2 =\pm(2\lambda^2 + \tfrac{N}{2} +o(1)),$$
 for certain constants $E$ and $N$.
\item Although they have different basepoints, we make the convention that the paths of integration for $\A(P)$ and $\Omega_j(P)$ are yoked,
so that a homology cycle may be added to one path only
if it is added to the other; in other words, the difference between the
paths for $\Omega_j(P)$ and $\A(P)$ is a fixed path from $\infty_-$ to $\lambda_{2g+2}$.
The most straightforward way of choosing this is to cut the Riemann surface along the
homology cycles $a_j,b_j$, resulting in a cut surface $\Sigma_0$ with boundary $\displaystyle\sum_j a_j + b_j - a_j - b_j$,
and fixing a path $\Pi$ between $\infty_-$ and $\lambda_{2g+2}$ in the interior of $\Sigma_0$ (see Figure \ref{cutting Sigma}).

\item Then {\sl frequency vectors} $\vV$ and $\vW$ are chosen to make $\psi^1,\psi^2$ 
well-defined functions on $\Sigma$. In fact, if the paths of integration are modified by adding an
integer linear combination $\sum_{k=1}^g n_k a_k + m_k b_k$ of homology cycles,
then $\vpsi$ changes by the factor
$$
\exp \left( \sum_{k=1}^g \left[ m_k \left( \ri s \oint_{b_k} \rd\Omega_1
+ \ri t \oint_{b_k} \rd\Omega_2 \right) -  m_k \left( \ri s V_k  + \ri t W_k\right)
\right] \right),
$$
which equals $1$ if we define the components of  $\vV$ and $\vW$ as
$$
V_k= \oint_{b_k} \rd\Omega_1, \qquad
\qquad W_k=\oint_{b_k} \rd\Omega_2 .
$$
\item ${\mathcal D}_\pm$ are the unique  positive divisors linearly equivalent to ${\mathcal D} -\infty_\pm$, and $\K$ is the vector of Riemann constants~\cite{Du}
which has the property that the zero divisor of $ \theta({\A}(P) - {\A}(\cD_\pm )-\K)$ is ${\mathcal D_\pm}.$
\item In order to make $\psi^1$ and $\psi^2$ have pole divisor  ${\cD}$,  the components of $\vpsi$ are  multiplied by meromorphic functions
$g_\pm$ whose zero divisors are ${\cD}_\pm +\infty_\mp$  and whose
poles lie in the original divisor ${\cD}$.
For the sake of definiteness, we will normalize
$$g_+(\infty_-) = g_-(\infty_+)=1.$$
\end{enumerate}

For later use (e.g., deriving formulas for the components
of the VFE solution curve $\gamma$ given by \eqref{generalSymformula}),
we will explain how to derive a more compact and convenient expression for the eigenfunction.
Let 
$$\vD = \A(\cD_+) + \K, \qquad \qquad \qquad  \vr = \int_{\infty_-}^{\infty_+} \omega,$$
both computed using paths in $\Sigma_0$.  Then $\A(\cD_-)+\K = \vD + \vr$, so
\begin{align*}
\psi^1 &= g_+(P) \exp\left(\ri s(\Omega_1(P)-\tfrac E2) + \ri t(\Omega_2(P)+\tfrac N2)\right)
 \dfrac{\theta(\A(P) + \ri \vV s +\ri \vW t -\vD)\,\theta(\vD)}
{\theta(\ri \vV s +\ri \vW t -\vD)\,\theta(\A(P)-\vD)},\\
\psi^2 &= g_-(P) \exp\left(\ri s(\Omega_1(P)+\tfrac E2) + \ri t(\Omega_2(P)-\tfrac N2)\right)
\dfrac{\theta(\A(P) + \ri \vV s +\ri \vW t -\vD-\vr)\,\theta(\vD)}{\theta(\ri \vV s +\ri \vW t -\vD)\,\theta(\A(P)-\vD-\vr)}.
\end{align*}

Next, introduce another meromorphic differential $\rd\Omega_3$, uniquely defined by
$$\rd\Omega_3 \sim \pm \dfrac{\rd\lambda}{\lambda}\qquad \text{as}\  P \to \infty_\pm,$$
and the requirement that it has zero $a$-periods.
Let $\Omega_3(P)$ be the corresponding Abelian integral on $\Sigma$,
computed with the same basepoint and convention for paths as $\Omega_1$,$\Omega_2$.
Then, by imposing reality conditions
that guarantee  that the VFE solution is a curve in $\mathbb R^3$ at all times (see \S\ref{realityconds} for details),
we obtain the following simplified expression for the components of $\vpsi$:
\begin{align}
\psi^1 &= \exp\left(\ri s(\Omega_1(P)-\tfrac E2) + \ri t(\Omega_2(P)+\tfrac N2)\right)
 \dfrac{\theta(\A(P) + \ri \vV s +\ri \vW t -\vD)}
{\theta(\ri \vV s +\ri \vW t -\vD)}, \label{psi1simple}\\
\psi^2 &= -\ri\exp\left(\ri s(\Omega_1(P)+\tfrac E2) + \ri t(\Omega_2(P)-\tfrac N2)+\Omega_3(P)\right)
\dfrac{\theta(\A(P) + \ri \vV s +\ri \vW t -\vD-\vr)}{\theta(\ri \vV s +\ri \vW t -\vD)}. \label{psi2simple}
\end{align}
An explicit computation involving the asymptotic expansions of the eigenfunction
$\vpsi$ at $\infty_\pm$ (see, e.g., \cite{BBEIM})
shows that $\vpsi$ is a common solution of the pair of NLS linear systems \eqref{LINS}
with potential $q$ given by
\begin{equation}
\label{qform in section 2}
q(x,t) = A \exp(-\ri E s +\ri N t)\dfrac{\theta(\ri \vV s+\ri \vW t-\vD+\vr)}{\theta(\ri \vV s + \ri \vW t -\vD)},
\end{equation}
where the formula for the constant $A$ is given in \S\ref{realityconds}.

\subsection{Curve Formulas}
\label{curveformula}

When $\pi(P)$ is real, a fundamental solution matrix of  (\ref{LINS})  is given by
\begin{equation}
\label{psipsibarmatrix}
\Psi(P; s,t)   = \left( \begin{array}{cc}
\psi^{1} (P) & -\overline{\psi^2(P)}\\ \psi^{2}(P) & \overline{\psi^1(P)}
\end{array} \right).
\end{equation}
The  Sym-Pohlmeyer reconstruction formula (\ref{RECO})
yields the following expression for the components of
the skew-hermitian matrix $\Gamma$ representing the position vector of the VFE solution:
\begin{equation}
\label{finalGammaformulas}
\begin{aligned}
\Gamma_{11} &= i(\Omega'_(P) s + \Omega'_2(P) t)
\\
&+\dfrac1{2 \det \Psi}\left[
|\psi^1|^2 \nabla \log\left(
\dfrac{\theta(\A(P)+  \phase)}{\theta(\A(P)-\phase)} \right)
+|\psi^2|^2 \nabla\log\left(
\dfrac{\theta(\A(P) +\phase-\vr)}{\theta(\A(P)-\phase-\vr)} \right)
\right] \cdot \dfrac{\rd\A(P)}{\rd\lambda},\\
\Gamma_{21} &=\dfrac{\psi^1\psi^2}{\det\Psi}\left(
\nabla\log\left(\dfrac{\theta(\A(P)+\phase-\vr)}{\theta(\A(P)+\phase)}\right)
\cdot\dfrac{\rd\A(P)}{\rd\lambda}+\Omega'_3(P)
\right),
\end{aligned}
\end{equation}
where $\phase= \ri \vV s +\ri \vW t -\vD$, and $\Omega'_j(P)=\rd\Omega_j(P)/\rd\lambda$.

It is easy to verify that, given a fundamental solution matrix $\Psi$ of the linear systems (\ref{LINS}), the quantity $\det{\Psi}=|\psi^1|^2+|\psi^2|^2$ is independent of both $s$ and $t$. Differentiating it with respect to $s$, one gets
\begin{multline*}
\left[
|\psi^1|^2 \nabla \log\left(
\dfrac{\theta(\A(P)+\phase)}{\theta(\A(P)-\phase)} \right)
+|\psi^2|^2 \nabla\log\left(
\dfrac{\theta(\A(P)+\phase -\vr)}{\theta(\A(P)-\phase-\vr)} \right)
 -2\det\Psi\nabla \log \theta(\phase)  \right] \cdot \vV =0.
\end{multline*}
Differentiation with respect to $t$ yields a similar identity with $\vV$ replaced by $\vW$. Since the two frequency vectors span $\mathbb{C}^2$, one concludes that the quantity between the square brackets must vanish and obtains the following

\begin{proposition} The components $(\gamma_1,\gamma_2, \gamma_3)$ of the position vector $\vgamma$ of an N-phase solution of the Vortex Filament Equation are given by the following expressions
\begin{align*}
\Gamma_{21}= \gamma_1+\ri \gamma_2 &  =\dfrac{\psi^1\psi^2}{\det\Psi}\left(
\nabla\log\left(\dfrac{\theta(\A(P)+\ri \vV s +\ri \vW t -\vD-\vr)}{\theta(\A(P)+\ri \vV s +\ri \vW t -\vD)}\right)
\cdot\dfrac{d\A(P)}{d\lambda}+\Omega'_3(P)\right), \\
\Gamma_{11}=\gamma_3 &= \ri [\Omega'_1(P) s + \Omega'_2 (P)t] +  \nabla \log \theta(\ri \vV s +\ri \vW t -\vD) \cdot  \dfrac{d\A(P)}{d\lambda},
\end{align*}
where $\pi (P)=\Lambda_0 \in \bR$ is the reconstruction point.
\end{proposition}

\subsection{Closure Conditions}\label{closure}
The potential $q$ associated with a closed curve
of length $L$ by the Hasimoto map is $L$-periodic, but only up
to multiplication by a unit modulus constant:
$$q(s+L)=e^{\ri\phi}q(s),\qquad\text{where\ } \phi = \int_0^L \tau\,ds.$$
However, a quasiperiodic potential like this is related to a genuinely periodic
NLS solution by the NLS symmetry mentioned in the
introduction:
$$\tilde q(s,t) = \re^{\ri(a s-a^2 t)} q(s-2a t,t).$$
(Obviously, one chooses $a=-\phi/L$ to make $\tilde q$ a periodic potential.)
This symmetry lifts to the NLS linear system:
$$\vpsi(s,t;\tilde q, \tilde \lambda) =
\exp(\tfrac{\ri}2(as-a^2 t)\sigma_3) \vpsi(s-2at,t;q,\lambda),
\qquad \text{where} \ \tilde\lambda=\lambda-\tfrac12 a.$$
(This transformation, for fixed time $t=0$, appears in Grinevich and Schmidt \cite{GSc}.)
It follows that the continuous spectrum of $\tilde q$ is the same as that
of $q$, but translated by $-a/2$, and the Sym formula \eqref{generalSymformula}
yields the same curves
at corresponding values of $\lambda$:
$$\Gamma \vert_{\lambda=\Lambda_0} = \tilde \Gamma\vert_{\tilde\lambda=\Lambda_0-\frac12 a}.$$

Thus, we can without loss of generality assume that a closed curve of length $L$
is generated from a potential $q(s,t)$ that is $L$-periodic
in the variable $s$. Note that, for finite-gap solutions, the value of the
frequency vector $\vV$ is unchanged under horizontal translation $\lambda_j \mapsto \lambda_j - \frac12 a$
of the branch points, but the phase constant $E$ changes
by $E \mapsto E - a$.  In fact, our assumption that $q$ is $L$-periodic implies that $E$ has been translated to zero and the entries of $\vV$ are rational multiples
of $2\pi/L$.  Therefore, if $V$ is the greatest common
divisor of the entries of $\vV$, then $q$ has period $T=2\pi/V$, and $L$ is
an integer multiple of $T$.

Since the natural curvatures (which are one half the real and imaginary parts of $q(s,t)$)
are periodic, the corresponding
natural frame vectors must also be periodic.  Since those
are obtained by conjugating a fixed basis of ${\mathfrak{su}}(2)$ by the fundamental matrix
$\Psi$, it follows that $\Psi$ must be $L$-periodic or antiperiodic.
For finite-gap
solutions satisfying the periodicity condition, this amounts to assuming that $\Omega_1(P)$ is an integer multiple
of $\pi/L$.

Examining the formulas \eqref{finalGammaformulas}
shows that our assumptions
so far guarantee that $\Gamma_{21}$ is periodic, and that $\Gamma_{11}$
is periodic if and only if $\rd\Omega_1(P)$ vanishes.  We summarize these closure
conditions as:

\begin{prop}[Closure]
\label{closureprop}
Suppose $q(s,t)$ is a finite-gap potential of period $2\pi/V$ in $s$.  Then the curve
reconstructed from $q$ using the Sym formula \eqref{generalSymformula} with
$\Lambda_0=\lambda(P)$ is
smoothly closed of length $L=2n\pi/V$ if and only if
(a) $\rd\Omega_1(P)=0$ and (b) $\exp(\ri L\Omega_1(P))=\pm 1$.
\end{prop}

The points which fulfil the closure condition (b) are discrete points of the Floquet spectrum associated with the potential $q$, which can be described in terms of 
{\it Floquet discriminant}. The Floquet discriminant of $q$ is defined as 
the trace of the transfer matrix of $\Psi$ over the interval $[0,L]$:
$$\Delta(q;\lambda) = \tr \left(\Psi(s,t; \lambda) ^{-1}\Psi(s+L,t;\lambda)\right).$$
(Because the transfer matrix only changes by conjugation when we shift in $s$ or $t$, $\Delta$
is independent of those variables.) Then, the Floquet spectrum is defined as the region 
$$
\sigma(q)=\{\lambda \in \mathbb{C} \, | \, \Delta(q;\lambda)\in \mathbb{R}, -2 \leq \Delta \leq 2\}.
$$
Points of the {\it continuous spectrum} of $q$ are those for which the eigenvalues of
the transfer matrix have unit modulus, and therefore $\Delta(q; \lambda)$ is real and between $2$ and $-2$; in particular, the real line is part of the continuous spectrum. Points of the $L$-periodic  {\it discrete spectrum} of $q$ are those for which the eigenvalues of the transfer matrix are $\pm 1$, equivalently 
$\Delta(q; \lambda)=\pm 2$.
Points of the discrete spectrum which are embedded in a continuous band of spectrum have to be critical points for the Floquet discriminant (i.e., $\rd \Delta/ \rd\lambda$ must vanish at such points);  thus, the conditions in Proposition \ref{closureprop} may be rephrased as requiring $\Lambda_0$ to be a real zero of the quasimomentum differential that is also a double point. These conditions coincide with those derived by Grinevich and Schmidt \cite{GSc} using different reasoning.

For finite-gap potentials, we may use the simplified eigenfunction formulas \eqref{psi1simple},\eqref{psi2simple} to
write the Floquet discriminant as
$$\Delta(q; \lambda) = 2 \cos (L\Omega_1(P)), \qquad \lambda=\pi(P).$$
If we have an explicit formula for the discriminant, we can use it to locate
the points of the discrete spectrum; see \S\ref{floquetd} for an example.
However, it is very difficult, in general, to determine how to choose the branch points
so that a point of the discrete spectrum coincides with a zero of the
quasimomentum differential.  Even in genus one, this corresponds to an implicit
equation involving elliptic integrals.

\subsection{Periodic Self-Intersection}\label{selfint}
Whether or not the filament closes smoothly,
it may intersect itself after one period $T$ of the potential. (For example, this happens
with elastic rod centerlines \cite{IS}.)  In this section,
we will calculate the general condition under which this happens.
In the expression for $\Gamma_{11}$ in \eqref{finalGammaformulas}, the only term which is not automatically $T$-periodic is the linear term $\ri \Omega'_1(P) s$. Therefore,  $\Gamma_{11}$ is $T$-periodic if and only if $\Omega'_1(P)=0$, i.e.,
the first condition of Prop. \ref{closureprop} above is satisfied.  So, we will assume this condition.
Next, $\Gamma_{21}$ is $T$-periodic if and only if
$$\psi^1\psi^2 = -\ri \exp(2\ri(\Omega_1(P) s + \Omega_2(P) t) + \Omega_3(P))
\dfrac{\theta(\A(P)+\phase)\theta(\A(P)+\phase-\vr)}{\theta(\phase)^2}$$
is periodic.  But that amounts to assuming that $\lambda(P)$ is a point for the $T$-periodic
 discrete spectrum, giving a smoothly closed filament of length $T$.
However, self-intersection may also occur if the last factor in $\Gamma_{21}$ vanishes
for some $s$-value:
\begin{equation}\label{selfintcond}
\left.\nabla\log\left(\dfrac{\theta(\A(P)+\phase-\vr)}{\theta(\A(P)+\phase)}\right)
\cdot\dfrac{d\A(P)}{d\lambda}\right|_{s=s_0}+\Omega'_3(P) = 0.
\end{equation}
(Note that this factor is also $T$-periodic in $s$.)
Then $\Gamma_{21}$ is not $T$-periodic, but is equal to zero at a succession of
$s$-values, and the $T$-periodicity of $\Gamma_{11}$ guarantees that the
filament returns to the same point in $\R^3$.

For finite-gap solutions, the first self-intersection condition $\Omega'_1(P)=0$ is polynomial in $\lambda$.
In general, the second condition \eqref{selfintcond} depends on $\lambda, s_0$ and also $t$, so that
this kind of self-intersection may only be present at one time.  However,
in genus one, a shift in time may be offset by a shift in $s_0$, so that
these self-intersections persist in time.  Furthermore, in genus one the derivatives
of the logarithms may be expressed in terms of Jacobi zeta functions, and we may choose $s_0$ and $t$ to give $\phase=0$. This yields a polynomial condition in $\lambda$.  We conjecture that an analogous
polynomial condition occurs in higher genus.

\section{Genus One Solutions}
\label{genusone}
\renewcommand{\Re}{{\sf Re}}
\renewcommand{\Im}{{\sf Im}}
\newcommand{\F}{{\mathcal F}}

In this section, we will work out the representation of solutions of the filament flow in detail for the
case when the genus is one, when $\Sigma$ is an elliptic curve given by the equation
\bel{ellipticeq}{
\mu^2 =\prod_{j=1}^{2g+2} (\lambda-\lambda_j)
=  (\lambda-\lambda_1)(\lambda-\lambda_2)(\lambda-\blamda_1)(\lambda-\blamda_2).
}
The Floquet spectrum of a generic $2$-phase solution is shown in Figure \ref{genus1spectra}(a).
In particular, the intersection points $\alpha_1$ and $\alpha_2$ of the complex bands of spectrum with the real axis
are the real zeros of the quasimomentum differential $\rd\Omega_1$ given in equations
\eqref{quasidifform},\eqref{dOmega1form} below.
As discussed earlier, closed curves are obtained by selecting the branch points $\lambda_j$ so that one
of the $\alpha_j$ is also a double point, and using $\Lambda_0=\alpha_j$ in the
reconstruction formula \eqref{generalSymformula}.

\subsection{Elastic Rods}
Filaments generated by genus one potentials are interesting objects from geometric and topological points of view,
because they are the centerlines for {\em Kirchhoff elastic rods}.  These are thin rods that are
critical for an energy that consists of one term for the isotropic bending energy
(i.e., a multiple of $\int \kappa^2 ds$, where $\kappa$ is the Frenet curvature of the centerline) and a
term for the twisting of the rod about its centerline.  Solutions of this physical variational problem
are exactly the solutions of a geometric variational problem, where the Lagrangian is of the form
\bel{geometricL}{
\F = \int\tfrac12 \kappa^2\, \rd s + l_1 \int \tau\, \rd s + l_2 \int \rd s.
}
(The values of the parameters $l_1, l_2$ depending both on the relative weight of the
bending and twisting energies and on the boundary conditions in the physical variational problem;
see \cite{IS} and \cite{LS96} for details.)
Regarding $l_1,l_2$ as Lagrange multipliers, we can
think of Kirchhoff elastic rod centerlines as extremizing the
bending energy subject to constant total torsion and constant length constraints.
These curves include classical Eulerian elastic curves (for $l_1=0$)
and free elastica (for $l_1=l_2=0$).

The identification between travelling wave solutions of the NLS and elastic rod centerlines was
first discovered by Kida \cite{Kida}, who sought solutions of the filament flow that moved by rigid
motion (i.e., rotation and/or translation) plus possibly a translation along the curve.  (As we will see below,
under the Hasimoto map these filaments correspond exactly to finite-gap NLS potentials of genus one.)
Kida obtained
parametrizations for such curves, in terms of elliptic integrals, showed that there exist smooth closed
curves of this type, and identified the planar curves of this type with Eulerian elastica.

\begin{figure}[ht]
\centering
\includegraphics[height=2in]{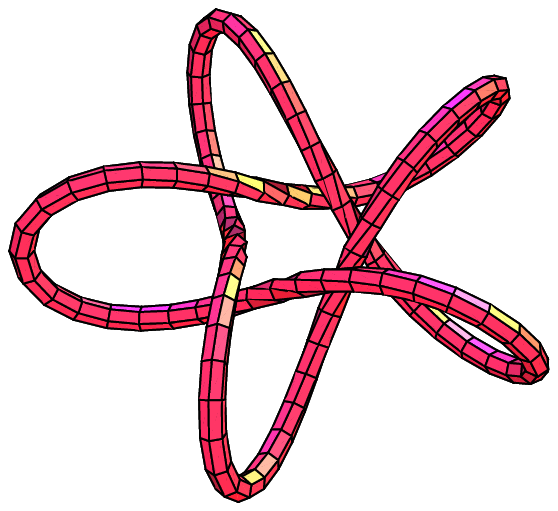}
\includegraphics[height=2in]{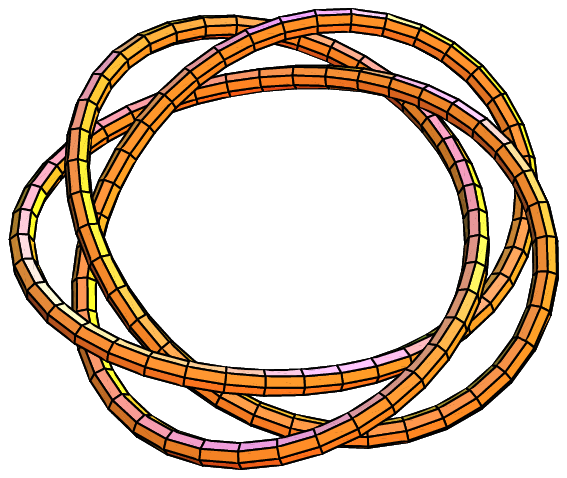}
\caption{Some examples of knotted elastic rods.}\label{trefoilrod}
\end{figure}

Later, Langer
and Singer \cite{LS96} began with the variational problem for $\F$ above, and showed that solutions
were precisely those curves which move by rigid motion (plus translation along the curve) under the filament flow.
Langer and Singer expressed the curvature and torsion of these curves in terms of periodic elliptic functions;
consequently, these curves consist of repeated congruent segments which smoothly close up
if certain elliptic integrals satisfy a rationality condition.
Figure \ref{trefoilrod} shows some examples.

Like these specimens, all closed elastic rod centerlines have discrete rotational symmetry about an axis.
It is not hard to see that if such a curve is knotted, it must be a torus knot.
Ivey and Singer~\cite{IS} later showed that every torus knot type is realized by a
one-parameter family of smooth elastic rod centerlines.  In fact, each such family connects
smoothly to another family of centerlines with the same rotational symmetry, but with
a complementary torus knot type (like $(2,5)$ and $(-3,5)$ shown above).   We call
this a {\em homotopy} of closed elastic rod centerlines.

\subsection{Genus One Potentials}\label{genus1integrals}
We now turn to the detailed calculation of genus one NLS potentials.
The data necessary to construct $q$ will be expressed in terms of elliptic integrals
and algebraic functions of the branch points.  For these calculations, we rely heavily on the treatise
by Byrd and Friedman \cite{BF}; citations of the form ``formula xxx.yy'' refer to this volume.

\begin{figure}[ht]
\centering
\includegraphics[height=2in]{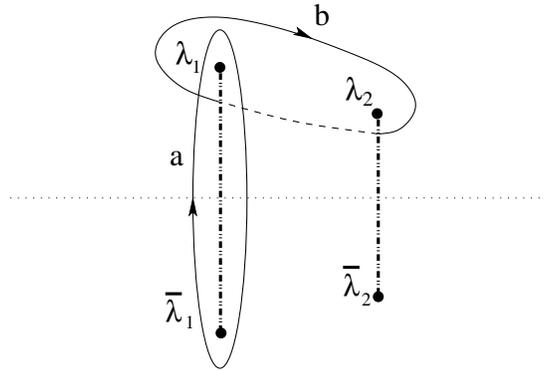}
\caption{Branch points for elliptic curve $\Sigma$ with oriented homology cycles and branch cuts.  The upper sheet tends to
$\infty_+$ to the right and left.}\label{g1basisfig}
\end{figure}

We begin by choosing a standard homology basis $a,b$ on the elliptic curve $\Sigma$ (see Figure \ref{g1basisfig})
and computing the periods of holomorphic differentials.

\begin{lem}\label{ablemma}
\bel{abperiods}{\oint_a \frac{\rd\lambda}\mu = \frac{4\ri K}{|\lambda_1-\blamda_2|},\qquad
\oint_b\frac{\rd\lambda}\mu = \frac{-4K'}{|\lambda_1-\blamda_2|},}
where we introduce the abbreviations
$$K := K(p),\qquad K':=K(p')$$
for complete elliptic integrals of the first kind with moduli
$$p':=\dfrac{|\lambda_1-\lambda_2|}{|\lambda_1-\blamda_2|},\qquad p:=\sqrt{1-(p')^2}.$$
\end{lem}
\begin{proof}  We will give a somewhat detailed explanation of this calculation, since
it is a model for others below.

To transform the integrals in \eqref{abperiods} to elliptic integrals
along the real axis, we use the linear fractional substitution
\bel{hformula}{\lambda(x)=\blamda_2+\dfrac{\lambda_2-\blamda_2}{1-h x},\qquad
h :=\dfrac{\lambda_1-\lambda_2}{\lambda_1-\blamda_2},}
under which $\lambda_1,\blamda_1,\lambda_2,\blamda_2$
correspond to $x$-values $1,(p')^{-2},0,\infty$ respectively.
Then \eqref{ellipticeq} gives
\bel{musqr}{\mu^2 = -h^2 |\lambda_1-\blamda_2|^2 \left(\dfrac{\lambda_2-\blamda_2}{1-h x}\right)^2
\left( x (x-1)(1-(p')^2 x) \right).}

Without changing
the value of the $a$-integral in \eqref{abperiods}, the $a$-cycle may be contracted in the $\lambda$-plane
to a straight-line path
that travels on the left-hand edge of the branch cut, from $\blamda_1$ to $\lambda_1$,
and then back along the right-hand edge of the branch cut.
Since $\mu$ differs only by a minus sign on the two sides of the branch cut, the $a$-integral is twice the integral along the
left-hand edge, where $\Re(\mu)>0$.
As $\lambda$ moves along this path, $x$ moves from $(p')^{-2}$ to $1$.
Thus $\Im(\rd\lambda)>0$ and $\Re(dy)<0$
as we move along this path, so the equation
$$\rd\lambda=\dfrac{h(\lambda_2-\blamda_2)}{1-hy}dy$$
shows that $\Re(h/(1-hy))<0$ there.  Hence, along this path we take
$$\mu = |(\lambda_1-\blamda_2)(\lambda_2-\blamda_2)|\left(\dfrac{-h }{1-hy}\right)\sqrt{x(x-1)(1-(p')^2x)},$$
giving
$$\frac12\oint_a \dfrac{\rd\lambda}{\mu}=\dfrac{-2\ri}{|\lambda_1-\blamda_2|}\int_{(p')^{-2}}^1
\dfrac{dy}{\sqrt{x(x-1)(1-(p')^2x)}}.$$
Reversing the limits in the integral on the right-hand side
yields the complete elliptic integral $K(p)$
(see formula 236.00).

The computation of the $b$-integral is similar.
\end{proof}

Consequently, the normalized holomorphic differential (with $a$-period equal to $2\pi \ri$)
is
$$\omega=2\pi \ri\left(\dfrac{ |\lambda_1-\blamda_2|}{4\ri K}\right) \dfrac{\rd\lambda}{\mu},$$
and the `Riemann matrix' for the $\theta$-function used in the
construction of $q$ is
\bel{Tgenus1}{B=\oint_b\omega=
-2\pi K'/K,}
The frequencies inside the $\theta$-function are
$$V=4\pi \ri\left(\oint_a \rd\lambda/\mu\right)^{-1}=\pi |\lambda_1-\blamda_2|/K$$
and $W=cV$, where
\bel{cformgenus1}{c=\sum_{j=1}^{2g+2} \lambda_j=\lambda_1+\blamda_1+\lambda_2+\blamda_2.}

Using the formulas \eqref{Eformula},\eqref{Nformula}, the frequencies in the exponential part of $q$ are given by
$$E=c-2L, \qquad N = 2cL - 4d - c^2,$$
where $c$ is as in \eqref{cformgenus1}, $d$ is given by \eqref{dformula}, and
$L$ is the following ratio of $a$-periods:
$$L := \left. \oint_a \dfrac{\lambda}{\mu} \rd\lambda \right/
\oint_a \dfrac{\rd\lambda}{\mu}.$$

To calculate the numerator of $L$, we use the same path and change of variables as in the proof
of Lemma \ref{ablemma}.  Thus, along the left edge of the branch cut, we have
$$\dfrac{\lambda}{\mu} \rd\lambda = \dfrac{-\ri}{|\lambda_1-\blamda_2|}
\left( \blamda_2 + \dfrac{\lambda_2-\blamda_2}{1-h x}\right)
\dfrac{dy}{\sqrt{x(x-1)(1-(p')^2 x)}}.$$
Applying formula 236.02, we get
\bel{Lform}{L =\blamda_2 + (\blamda_1- \blamda_2) \dfrac{\Pi( \beta^2, p)}K,}
where $\Pi(\beta^2,p)$ denotes a complete elliptic integral of the third kind, with parameter
$$\beta^2 := \dfrac{p^2}{1-\overline{h}} = \dfrac{\lambda_1 - \blamda_1}{\lambda_1 -\blamda_2}.$$
Note that we may also calculate $L$ using formula 235.02, obtaining
\bel{Lformalt}{L=\lambda_2+ (\lambda_1-\lambda_2)\dfrac{\Pi(\overline\beta^2,p)}K.}
This is equivalent to \eqref{Lform} by application of the addition formula 117.03 for elliptic integrals
of the third kind.  Furthermore, comparing \eqref{Lform} and \eqref{Lformalt}
shows that $L$ is real, as it must be since complex conjugation
$$\tau:(\lambda,\mu) \mapsto (\blamda,\widebar{\mu})$$
takes cycle $a$ to $-a$.

To calculate the shift $r$ in \eqref{qform in section 2}, we use the Abelian differential
$$\rd\Omega_3 = \dfrac{\lambda-L}{\mu}\rd\lambda$$
which has $a$-period equal to zero.
Using formula 233.02 to compute $\oint_b (\lambda/\mu) \rd\lambda$ gives
\bel{firstrformula}{r=-\oint_b \rd\Omega_3 = \dfrac{4}{|\lambda_1-\blamda_2|}
\left[(\blamda_2-L) K'+(\lambda_2-\blamda_2)\Pi(h,p')\right].}
One can verify that $\Im(r)=\pi$ by using the addition formula 117.02 to
simplify $\Pi(h,p)+\Pi(\overline{h},p)$, given that $\overline{h}=(p')^2/h$.

It will later be convenient to have an alternate formula for $r$.  Using addition formula 117.05, we can
write
$$(1-h)K \left(\Pi(h,p') - K'\right)
= h K' \Pi(\overline\beta^2,p) - \dfrac{\pi}2\sqrt{\dfrac{1-h}{\overline{h}-1}}F(\phi,p'),
\qquad \phi = \arcsin\left(\sqrt{h}/p'\right),$$
where $F$ denotes an incomplete elliptic integral of the first kind, e.g., $F(\pi/2,p')=K'$.
(Note that $|h|=(p')^2 <1$,
$\arg(h) \in [0,\pi)$, $\arg\left((1-h)/(\overline{h}-1)\right) = 2\arg(\lambda_2 - \blamda_1) \in (0,\pi)$,
and we take the square root to have positive real part.)
We use this formula to substitute for $\Pi(h,p')$ in \eqref{firstrformula}, and use \eqref{Lformalt}
to substitute for $L$, to get
\bel{secondrformula}{r = \dfrac{2\pi}{K} F(\phi,p')}

Finally, the modulus of scalar $A$ may be calculated by numerical integration as described in
Appendix \ref{practicalc}.  However,
we also have the following formula,
\bel{modAgenus1}{
|A| = \dfrac{|\lambda_1-\blamda_2|}{\theta(r)}\sqrt{\dfrac{2pp' K}{\pi}},}
which may be derived by substituting the formula \eqref{qform in section 2}
for the potential into the NLS and evaluating at $s=t=0$.

\subsection{Geometric Interpretation}
The above information is enough to enable us to discuss the geometric
invariants (i.e., curvature and torsion) for the filaments corresponding
to a genus one potential $q(s,t)$.

Because a translation in time modifies
$q$ only by a unit modulus factor which is independent of $s$, and may be
absorbed inside the theta function by translation in $s$, the geometric
invariants at time $t$ differ from those at time zero only by a constant-speed translation
in $s$.  Thus, the filament at time $t$ is obtained from the filament
at time zero by Euclidean motions
(accompanied by a translation along the curve).  It follows from the semigroup
property of the evolution equation that the Euclidean motions must belong to
a one-parameter subgroup.  Together with the discussion at the beginning of this
section, this gives the following
\begin{prop}{All finite-gap solutions of genus one for the vortex filament equation
move by rigid motions, and are congruent to elastic rod centerlines.}
\end{prop}

It is instructive to express the curvature of genus one solutions in terms of elliptic functions.
By the above argument, we may assume that $t=0$ without loss of
generality. Also, since the reality condition (see Appendix \ref{realityconds})
implies that $D$ is purely imaginary, we may assume that $D=0$.

We first observe that for $B$ given by \eqref{Tgenus1},
$$\theta(2\ri z) = \theta_3(z)=\theta_0(z + \pi/2),$$
where $\theta_k$ denotes the Jacobi theta function with
modulus $p$ and period $2\pi$ (for $k=1,2$) or $\pi$ (for $k=0,3$).
Next, let $\rho = \frac12\Re(r)$.  Then
\bel{qformsimple}{q(s,0) = A \exp(-\ri E s) \dfrac{\theta_3( (V s+\pi)/2 - \ri\rho)}{\theta_0((V s+\pi)/2).}}
For convenience, introduced the rescaled variable $z=(Vx+\pi)/2$.  Then,
using the translation properties and addition formulas 1051.ff for
Jacobi theta functions, we have
$$
\kappa^2 = |2q(s,0)|^2 = |2A|^2\dfrac{ \theta_3(z +\ri\rho)\theta_3(z-\ri\rho)}{\theta_0^2(z)}
= |2A|^2 \dfrac{\theta_3^2(\ri\rho)}{\theta_0^2(0)}
    \left(1 - \dfrac{\theta_2^2(\ri\rho)\theta_1^2(z)}{\theta_3^2(\ri\rho)\theta_0^2(z)}\right).
$$
This formula coincides with the
curvature formula for elastic rod centerlines derived in \cite{LS96}; for, if we set
\bel{wdef}{w := \dc(2Ki\rho/\pi)=\dn(2K\rho/\pi,p'),} and substitute
in for $|A|$ from \eqref{modAgenus1}, then we have $$\kappa^2 = \kappa_0^2
\left( 1 - \dfrac{p^2}{w^2} \sn^2 u\right), \qquad \text{where}\qquad
u = \dfrac{2K}\pi z = \dfrac{\kappa_0}{2w}(s+\pi/V),$$ and we use
$\kappa_0 = 2w|\lambda_1 - \blamda_2|$ to denote the maximum
curvature.

\bigskip
We will now make some specific observations about the relation between
the geometry of the filament and the spectrum in genus one.
For these, we will need the following formula:
\begin{lem}
\bel{wformula}{w^2 = \dfrac{|\Im(\lambda_1 -\blamda_2)|^2}{|\lambda_1- \blamda_2|^2}}
\end{lem}
\begin{proof} Using \eqref{secondrformula}, we compute that
$$w = \dn(2K\rho/\pi, p') = \ri \cs(Kr/\pi,p') = \ri \cn(2F,p')/\sn(2F,p'),$$
where $F=F(\phi,p')$, as in \eqref{secondrformula}.
Then, using double angle formulas 124.01,
$$w=\dfrac{\sn^2(F,p')\dn^2(F,p') - \cn^2(F,p')}{2\ri\sn(F,p') \cn(F,p') \dn(F,p')}
=\pm\dfrac{\Re(h-1)}{|h-1|},$$
where the ambiguity of sign arises from solving $\sn(F,p')=\sqrt{h}/p'$ for the values
of $\cn(F,p')$ and $\dn(F,p')$. Then
substituting \eqref{hformula} yields the result.
\end{proof}

In \cite{LS96}, it is shown that the curvature and torsion of an
elastic rod centerline satisfy
\bel{taufromLS}{
2\tau - l_1 = j/\kappa^2,\quad\text{where}\quad j^2 = \dfrac{\kappa_0^6}{w^4}(1-w^2)(w^2-p^2),}
and $l_1$ is the ratio of the coefficient of the term $\int \tau\,ds$ to the
coefficient of the elastic energy $\int \tfrac12 \kappa^2\,ds$ in the
Lagrangian \eqref{geometricL}.
It is clear that the torsion of the centerline is
constant if and only if $w^2=p^2$ or $w^2=1$.  From \eqref{wformula} we see
that these correspond to the cases where $\lambda_1-\lambda_2$
is purely real or purely imaginary, respectively.  In these cases,
we may translate the branch points so that they are symmetric under the
reflection $\lambda \mapsto -\lambda$ about the origin.  (Recall that the
curve produced by the Sym-Pohlmeyer formula is unchanged provided we apply the
same translation to the reconstruction point $\Lambda_0$.)  We summarize these
observations:

\begin{prop}  A filament associated to a finite-gap solution of genus one
has constant torsion if and only the branch points are symmetric about
an axis perpendicular to the real axis.
\end{prop}
 In \S\ref{symmetric} of this paper, we will discuss what
symmetry of the branch points means in higher genus.

\subsection{Quasimomentum Differential}
Before making further observations, we need to calculate the meromorphic differential $\rd\Omega_1$ for genus one.
(Recall that the zeros $\alpha_k$ of this differential are used to reconstruct closed curves.)  This
will later enable us to calculate the Floquet discriminant, and
(in Prop. \ref{alphacoalesce} below) to connect the coalescence of the $\alpha_k$'s
with the filament being an elastic curve.

The form of this differential is
\bel{quasidifform}{
\rd\Omega_1 = \dfrac{\lambda^2 - \tfrac{c}2 \lambda - c_1}{\mu}\, \rd\lambda,
}
where $c$ is as in \eqref{cformgenus1} and $c_1$ is chosen so that the $a$-period
of $\rd\Omega_1$ is zero.  The value of $\oint_a (\lambda/\mu)\rd\lambda$ was obtained
in the above calculation of $L$.  In a similar way,
we apply formulas 236.17 and 336.02 to get
\begin{multline}
\oint_a\dfrac{\lambda^2}{\mu} \rd\lambda =\dfrac{4\ri}{|\lambda_1-\blamda_2|}\left[ \blamda_2^2 K
+ 2 \blamda_2 (\blamda_1 -\blamda_2) \Pi(\beta^2,p)\right.\\
+\left.\dfrac{(\blamda_1 -\blamda_2)^2}{2(1-\beta^2)(\beta^2 -p^2)}
\left(\beta^2 E(p) + (p^2-\beta^2)K + (2\beta^2 p^2 +2\beta^2 -\beta^4 - 3p^2)\Pi(\beta^2,p)\right)
\right],
\end{multline}
where $E(p)$ is the complete elliptic integral of the second kind.
We combine the previous two results to compute
$$c_1 = \left. \oint_a \dfrac{\lambda^2 - \tfrac{c}2 \lambda}{\mu} \rd\lambda \right/
\oint_a \dfrac{\rd\lambda}{\mu}.$$
Miraculously, the coefficient of $\Pi(\beta^2,p)$ in $c_1$ is zero,
giving
\bel{dOmega1form}{\rd\Omega_1 = \left[\lambda^2 - \tfrac{c}2 \lambda -
\tfrac12 \left( |\lambda_1-\blamda_2|^2 \dfrac{E(p)}{K} -
|\lambda_1|^2 - |\lambda_2|^2\right) \right] \dfrac{\rd\lambda}{\mu}.}

The roots of the quadratic polynomial in $\lambda$
 enclosed in square brackets are the zeros $\alpha_1,\alpha_2$ of the
quasimomentum differential.
\begin{prop}\label{alphacoalesce} Suppose a genus one filament is a quasiperiodic space curve, i.e.,
it has periodic curvature and torsion, and is bounded in space.  Then
the filament is an elastic curve if and only if $\alpha_1=\alpha_2$.
\end{prop}
\begin{proof}
In \cite{LS96} (see equation (26) in that paper), it is shown that elastic rod centerlines are bounded in space if and only if
$$\dfrac{2E(p)}{K} - 1 = w^2 - p^2+\dfrac{w^2 l_1^2}{\kappa_0^2},$$
where $l_1$ is the coefficient in the geometric Lagrangian \eqref{geometricL} for elastic rod centerlines.
Thus, quasiperiodic elastic rod centerlines are elastic curves if and only if
\bel{EKform}{\dfrac{2E(p)}{K} - 1 = w^2 - p^2}
On the other hand, the discriminant of the quadratic in \eqref{dOmega1form} is
$$\dfrac{c^2}4 + 4c_1 = \tfrac14(\lambda_1 + \blamda_1+ \lambda_2 + \blamda_2)^2+
2|\lambda_1-\blamda_2|^2\dfrac{E(p)}{K}- 2|\lambda_1|^2 - 2|\lambda_2|^2.$$
By using \eqref{wformula}, one sees that this vanishes if and only if \eqref{EKform} holds.
\end{proof}

\begin{remark} The assumption of quasiperiodicity seems to be necessary here.  For, without changing
the spectrum of the curve, we can shift the reconstruction point $\Lambda_0$ in the Sym-Pohlmeyer formula; this will
add a constant to the torsion, causing the curve to no longer be quasiperiodic, and also causing
the curve to cease to be an elastic curve.
\end{remark}

\subsection{Floquet Discriminant}\label{floquetd}
Following our earlier discussion
the trace of the transfer matrix over one period of $q$ will be given by
$$\Delta(\lambda) = 2 \cos \left(\dfrac{2\pi}V \Omega_1(P)\right),$$
where $P$ is a point on the hyperelliptic curve lying over $\lambda$ and $\Omega_1(P)$ is
the integral of $\rd\Omega_1$ from basepoint $\blamda_2$ to $P$.  This integral is path-dependent,
but since $\rd\Omega_1$ has zero $a$-period and $b$-period equal to $V$, the right-hand side is unaffected by the choice
of path.  It is also unaffected by the choice of branch for $P$, since $\interchange^*\rd\Omega_1=-\rd\Omega_1$, where $\interchange$ is the sheet interchange involution.

Because the integral formulas used
in \S\ref{genus1integrals} are better suited to using basepoint $\blamda_1$, we will do that instead.
Because $\int_{\blamda_1}^{\lambda_1}\rd\Omega_1 = \int_{\blamda_2}^{\lambda_2}\rd\Omega_1=0$
and $\int_{\lambda_1}^{\lambda_2}\rd\Omega_1 = \pm V/2$, this
change of basepoint changes the value
of $\Delta(\lambda)$ by a minus sign.

Let $\Gamma$ denote the path chosen from $\blamda_1$ to $P$.
Initially, we will assume that $P$ and $\Gamma$ lie on the upper sheet, to
the left of the branch cut from $\blamda_1$ to $\lambda_1$.  (On this sheet,
$\Re(\mu)>0$ above the real axis.)  However, the formula for $\Delta(\lambda)$
which we will derive extends analytically for all $\lambda$.

Let $X$ denote the $x$-value corresponding to $\lambda$ under the change of variables.  Then
by a calculation similar to the proof of Lemma \ref{ablemma} (in particular, using 236.00 in \cite{BF}),
$$\int_\Gamma \dfrac{\rd\lambda}\mu = \dfrac{-\ri}{|\lambda_1-\blamda_2|}\int_{(p')^{-2}}^{X}
\dfrac{dy}{\sqrt{x(x-1)(1-(p')^2x)}} =
\dfrac{2\ri}{|\lambda_1-\blamda_2|} F(\psi,p),
$$
where
$$\psi = \arcsin\left(\sqrt{1-X (p')^2}/p\right).$$

Let $u=F(\psi,p)$.  Note that this is an analytic function of $\psi$ provided
$|\Re(\sin\psi)|<1/p$.  In fact, this is true if we restrict $X$ to
lie strictly inside the circle in the $x$-plane centered on the real axis
and passing through $1$ and $(p')^{-2}$.  Then we may take
\bel{snoo}{\sn u = \sqrt{1-X (p')^2}/p,
\qquad\cn u = \dfrac{p'}{p}\sqrt{X-1},\qquad \dn u = p' \sqrt{X}.}

By a calculation similar to that yielding $L$, we get
$$\int_\Gamma \dfrac{\lambda }\mu \rd\lambda = \dfrac{2\ri}{|\lambda_1-\blamda_2|}
\left(\blamda_2 u + \dfrac{\lambda_2-\blamda_2}{1-h(p')^{-2}}\Pi(\psi,\beta^2,p)\right),$$
where $\Pi(\psi,\beta^2,p)$ denotes an incomplete elliptic integral of the third kind.
However, in calculating the $\Gamma$-integral of the quasimomentum differential $\rd\Omega_1$,
the coefficients of $\Pi(\psi,\beta^2,p)$ cancel out, so we will omit these terms from
now on.

Applying formulas 236.17 and 336.02 again, we get
\begin{multline}
\int_\Gamma\dfrac{\lambda^2}{\mu} \rd\lambda =\\ \dfrac{2\ri}{|\lambda_1-\blamda_2|}\left[
\blamda_2^2 u
+\dfrac{(\blamda_1 -\blamda_2)^2}{2(1-\beta^2)(\beta^2 -p^2)}
\left(\beta^2 E(\psi,p) + (p^2-\beta^2)u -\beta^4\dfrac{\cn u \dn u \sn u}{1-\beta^2\sn^2 u}\right)\right].
\end{multline}
Finally, combining these terms gives
\begin{align*}\int_\Gamma \rd\Omega_1
&= \int_\Gamma \dfrac{\lambda^2 - \tfrac{c}2 \lambda - c_1}{\mu}\, \rd\lambda\\
&= \ri|\lambda_1-\blamda_2|\left( Z(u) - \beta^2 \dfrac{\cn u \dn u \sn u}{1-\beta^2\sn^2 u}\right),
\end{align*}
where $Z(u)$ denotes the Jacobi zeta function with modulus $p$.  Consequently, the Floquet
discriminant is
\bel{discform}{
\Delta(\lambda)=-2\cos\left[2\ri K\left( Z(u) - \beta^2 \dfrac{\cn u \dn u \sn u}{1-\beta^2\sn^2 u}\right)\right],
\qquad u = F\left(\arcsin\left(\sqrt{1-X (p')^2}/p\right)\right).}
(The minus sign in front arises from our change of basepoint from $\blamda_2$ to $\blamda_1$.)

In an earlier paper \cite{CIsg}, we made a similar calculation of the discriminant for genus one solutions of the
sine-Gordon equation.  Like that formula, the argument of the cosine in $\Delta(\lambda)$ here
consists of a Jacobi zeta function minus an algebraic term.
For, by substituting in from \eqref{snoo} and comparing with \eqref{musqr}, we obtain
$$\beta^2 \dfrac{\cn u \dn u \sn u}{(1-\beta^2\sn^2 u)} =
\dfrac{\ri}{|\lambda_1-\blamda_2|}\left(\dfrac{\mu}{\lambda-\blamda_2}\right),
$$
where sign may be checked by verifying that the left-hand side has
positive real part when, say, $X=1/p'$.  (Recall that we are assuming that $\Re(\mu)>0$.)
For the purposes of investigating the discriminant numerically, this alternate expression
is not useful because the branch cut involved in calculating $\psi$ does not necessarily
coincide with the branch cut for $\mu$.  However, the simultaneous branch cuts in the
two terms in \eqref{discform} cancel out, modulo $2\pi$, showing that $\Delta$ is a smooth function of $\lambda$.

The spectrum may be plotted using level curves of
$\realpart(\Delta(\lambda)$ and $\Im(\Delta(\lambda))$; for, the spectrum
is contained in the level curve  $\Im(\Delta(\lambda))=0$, the simple points (which
lie at the end of the continuous spectrum) lie at the intersection of this
level curve with a level curve where $\realpart(\Delta(\lambda)=\pm2$,
and  points of order $k$ lie at the intersection of $k$ such level curves with
the continuous spectrum.

\setlength{\tricklength}{.1in}
\begin{figure}[ht]
\centering
\begin{tabular}{c@{\hspace{.2in}}c}
\includegraphics[height=1.75in]{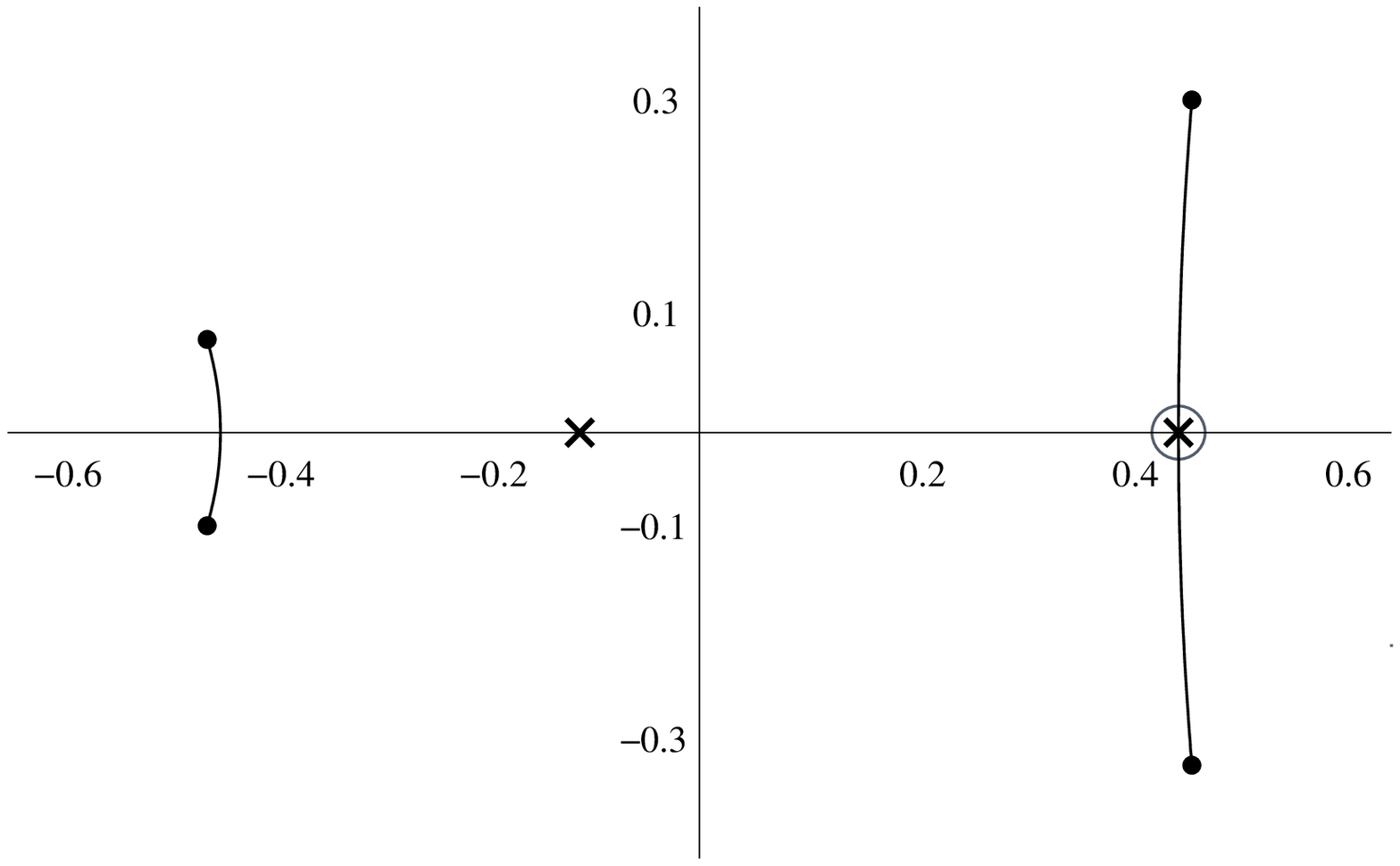}&
\includegraphics[height=1.75in]{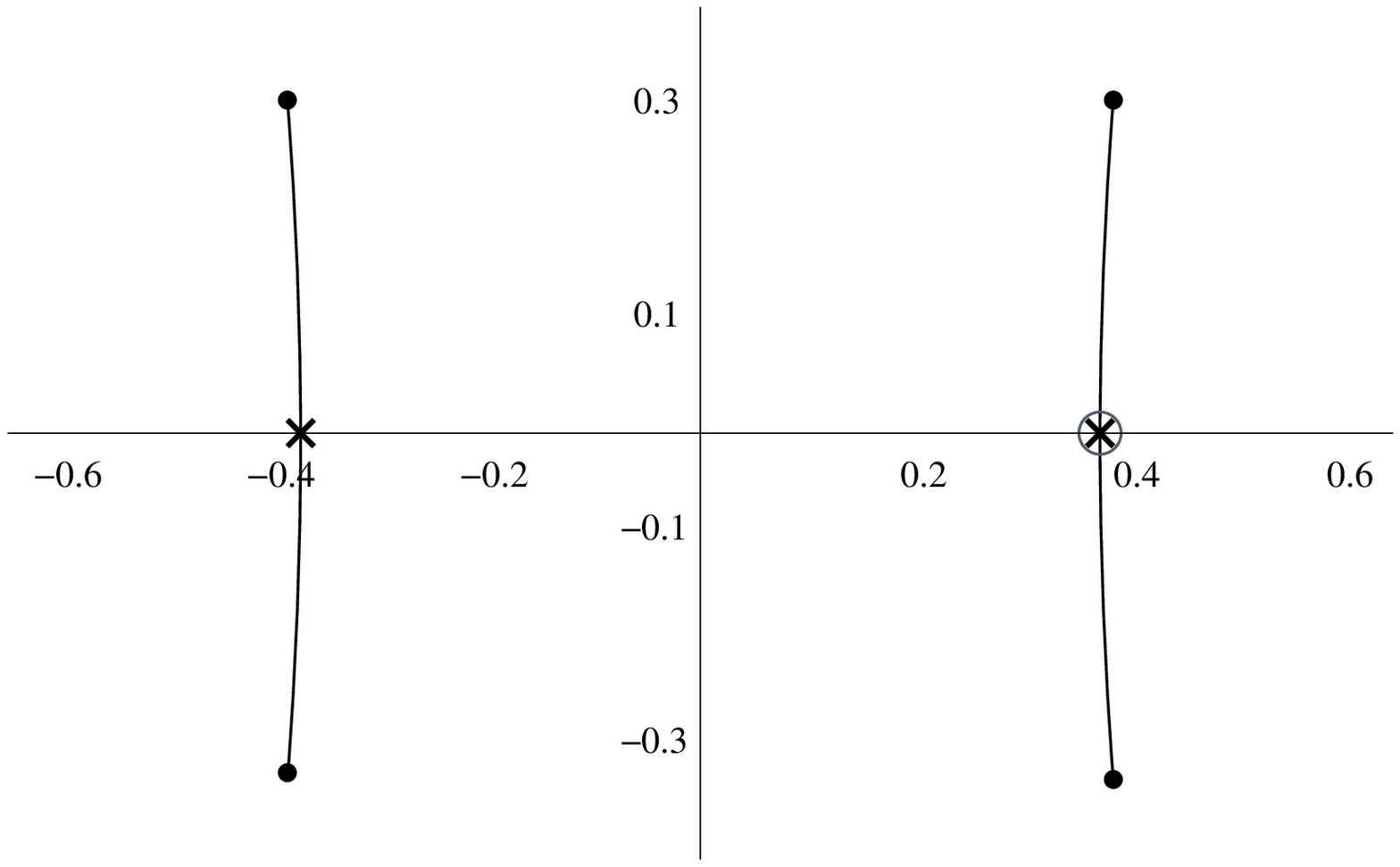}\\[-\tricklength]
{\bf(a)}& {\bf(b)}\\[\tricklength]
\includegraphics[height=2.5in]{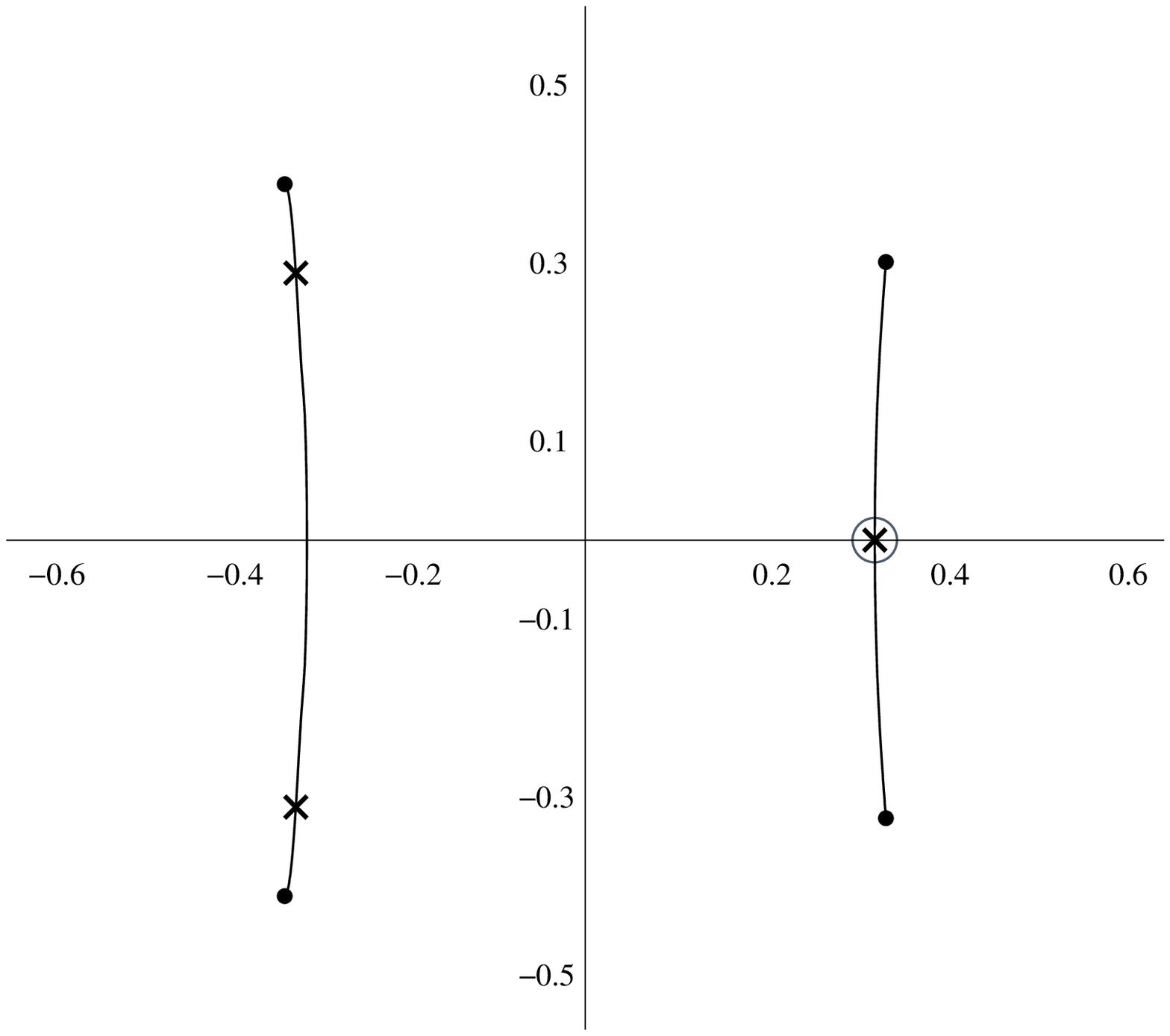}&
\includegraphics[height=2.5in]{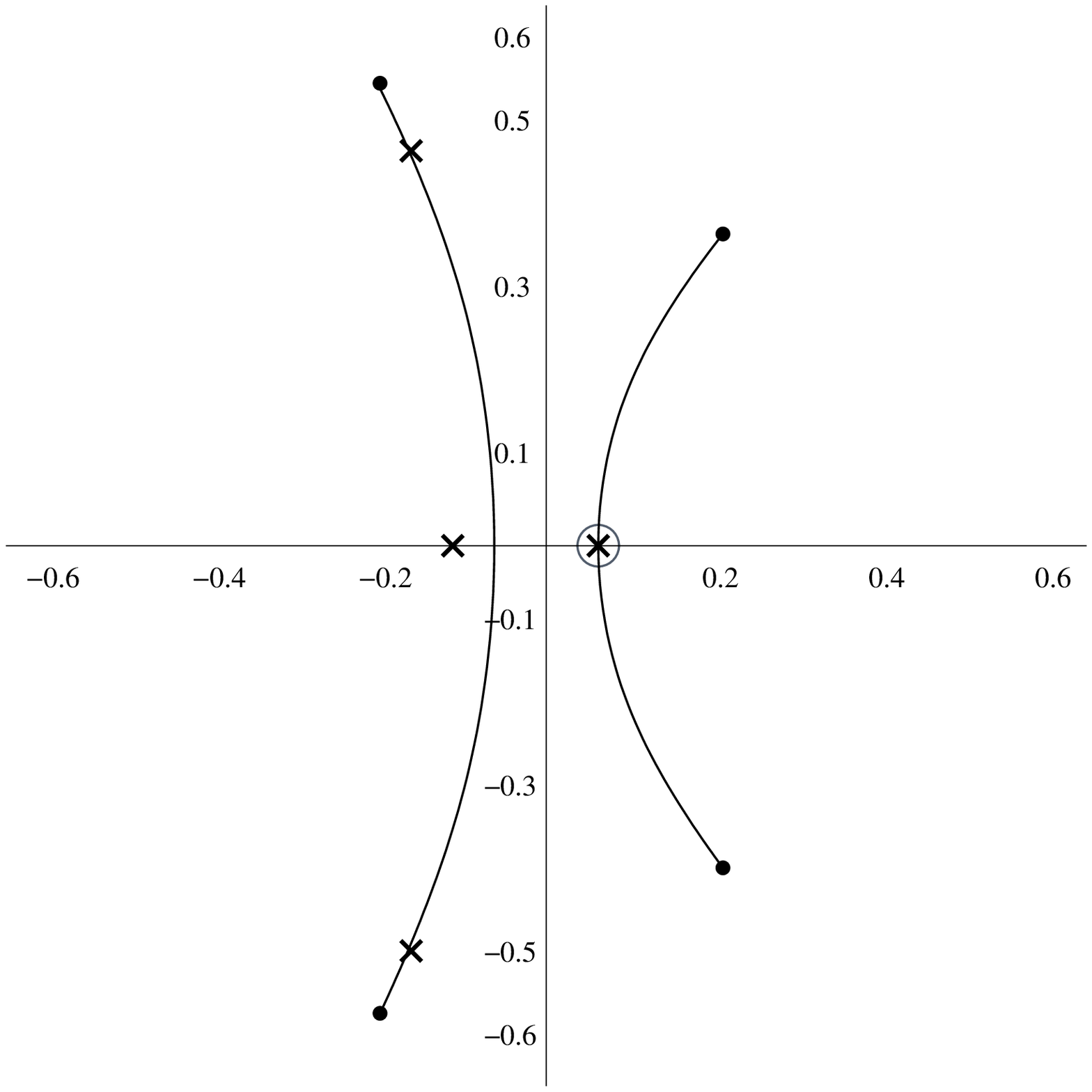}\\[-\tricklength]
{\bf(c)}& {\bf(d)}\\[\tricklength]
\includegraphics[height=2.5in]{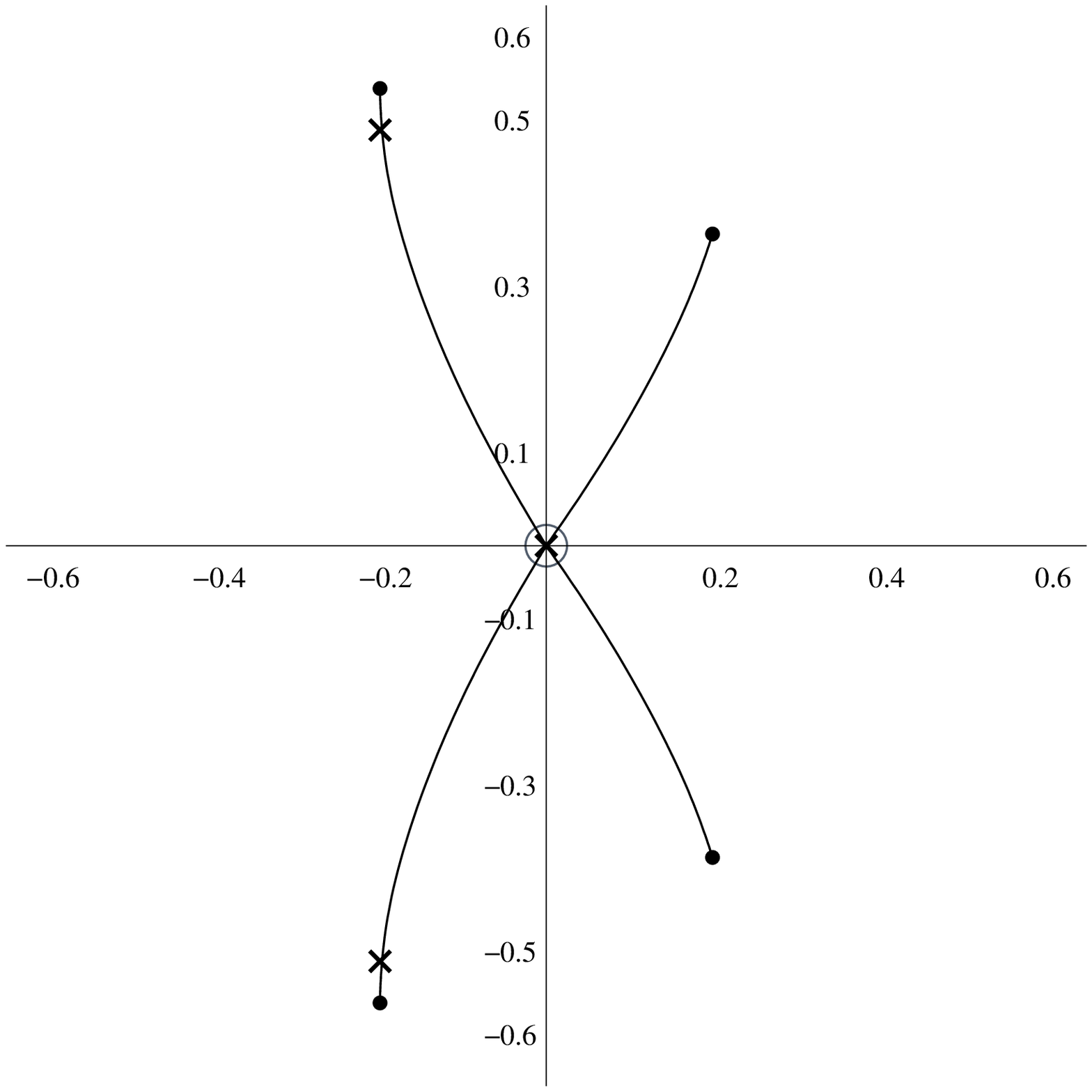}&
\includegraphics[height=2.5in]{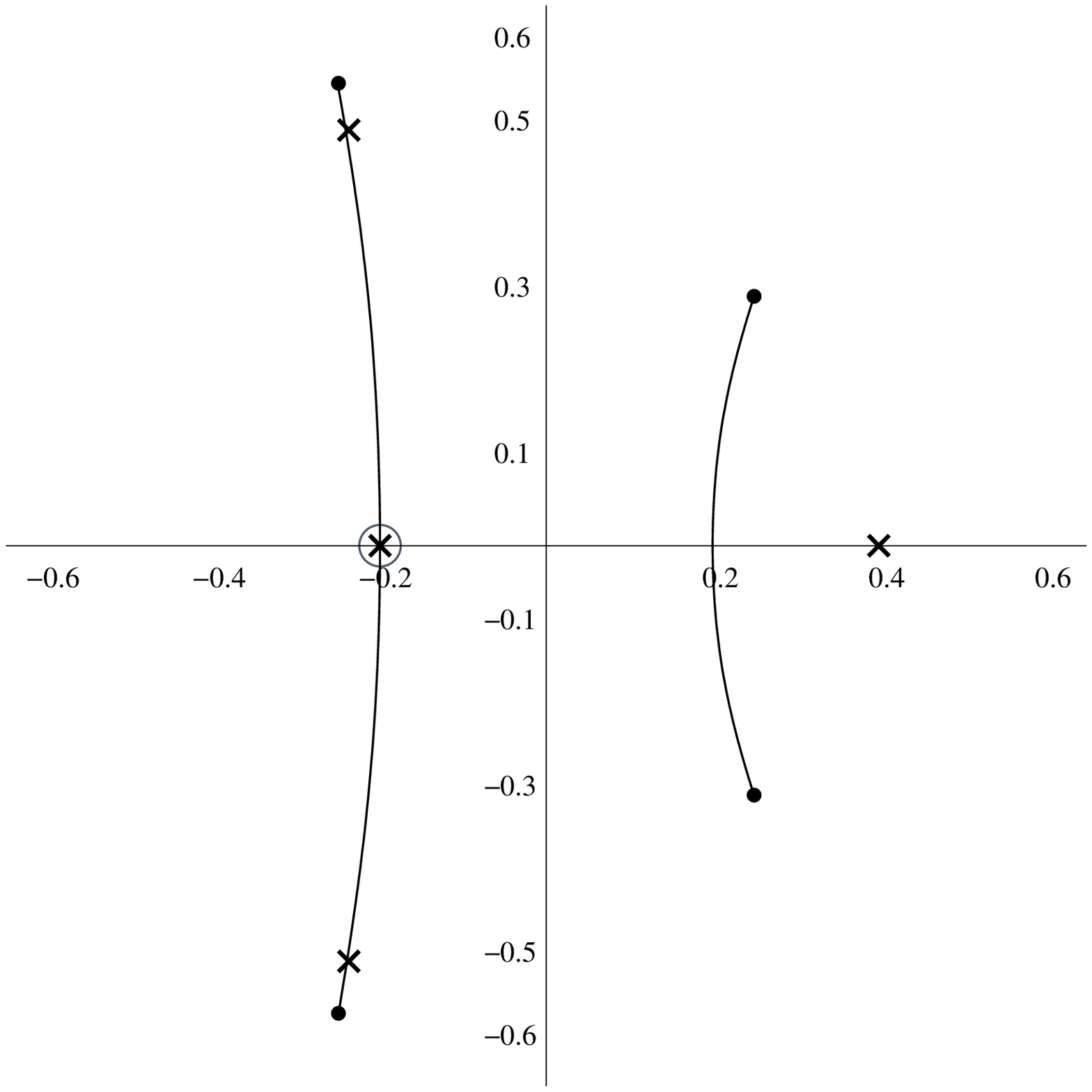}\\[-\tricklength]
{\bf(e)}& {\bf(f)}\\[\tricklength]
\end{tabular}
\caption{Drawings of the spectra of selected closed elastic rods belonging
to a homotopy between a singly-covered circle and a double-covered circle,
going from unknotted to a trefoil knot type
along the way.}\label{genus1spectra}
\end{figure}

\subsubsection*{About Figure \ref{genus1spectra}}
This figure shows the Floquet spectra of some elastic rods (including some with
special geometric features) belonging to a typical homotopy.
The diagrams, which are based on Mathematica plots using equation \eqref{discform},
show the continuous spectrum, including the real axis, the simple points (which
appear as dots), the multiple points (which are usually double points,
and appear as x's), and the Sym-Pohlmeyer reconstruction point (which is circled).\footnote{%
When the curve is smoothly closed, this point is necessarily a multiple point of order at least four \cite{CI3}.}
In each case, the spectrum is translated in the real direction so that the sum of the
branch points is zero.
The stages of the homotopy shown are:
\begin{enumerate}
\item[(a)]
an unknotted rod, with branch points $-0.454\pm 0.095\ri$ and $0.454\pm 0.324\ri$
in the upper half plane;
\item[(b)] a constant torsion rod, with symmetric spectrum and
branch points $\pm 0.388 \pm 0.316 \ri$ (note the coalescence,
at the base of the left-hand spine,
of two double points, one of which previously entered from the left);
\item[(c)] another unknotted rod, with branch points
$-0.342 \pm 0.409846 \ri$ and $0.342 \pm  0.320\ri$, and double
points migrating out along the left-hand spine;
\item[(d)] a third unknotted rod, with branch points $-0.206 \pm 0.545\ri$ and
$0.205 \pm 0.366\ri$, and a double point which has entered from the left;
\item[(e)] an Eulerian elastic curve, with branch points
$-0.201\pm 0.546\ri$ and $0.201\pm 0.370\ri$, and a multiple point
of order six at the origin, where the zeros of the quasimomentum differential have coalesced;
\item[(f)] a knotted rod, with branch points $-0.251\pm 0.557\ri$ and $0.251\pm 0.308\ri$,
and the reconstruction point now on the left-hand spine.
\end{enumerate}
As the homotopy continues, the right-hand spine shrinks toward the real axis.
After the change in knot type, the spectrum does not
look qualitatively different from frame (f).

\setlength{\tricklength}{.1in}
\begin{figure}[ht]
\centering
\includegraphics[height=.9\textheight, trim =80 220 80 10]{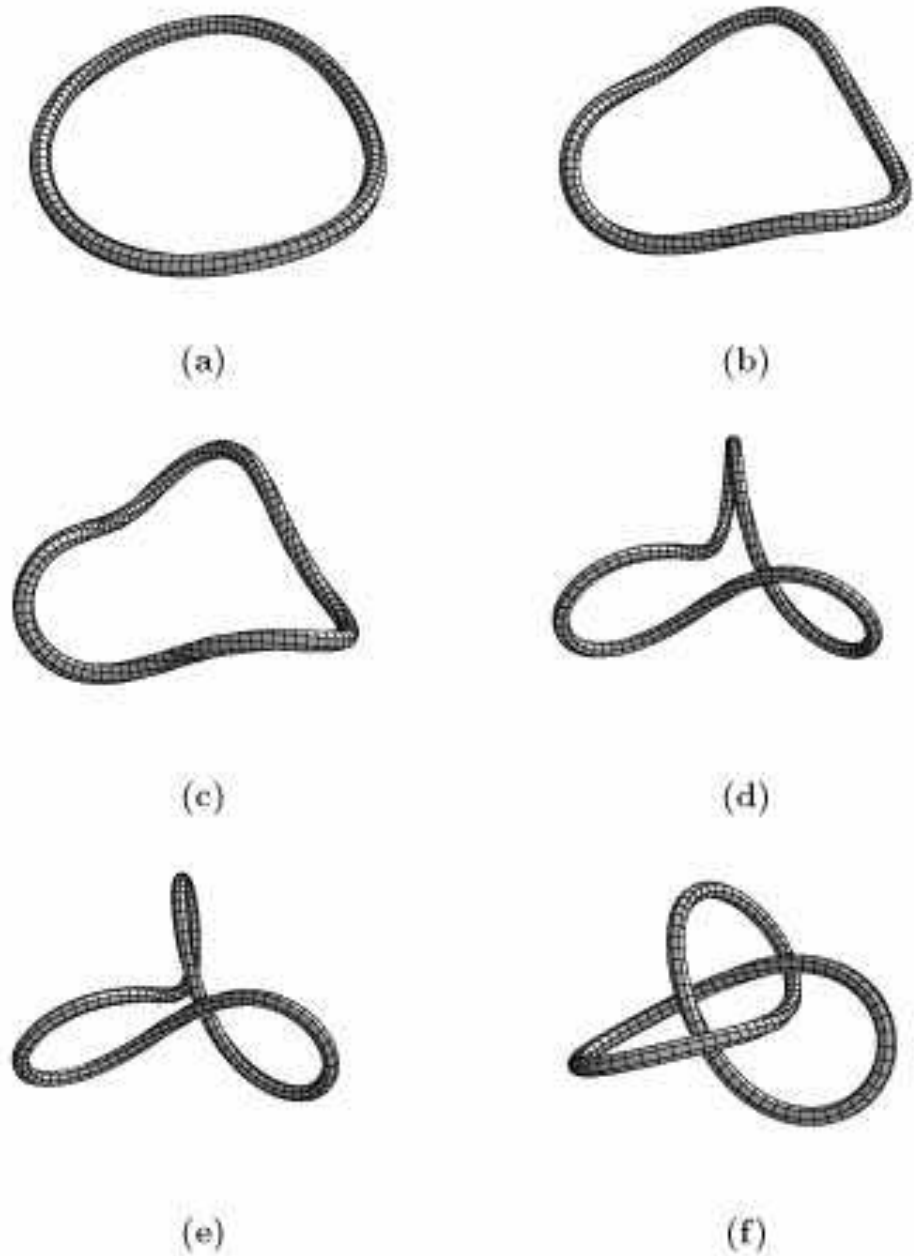}
\caption{Drawings of the the rod centerlines corresponding to
the spectra in Figure \ref{genus1spectra}.}
\end{figure}

\clearpage

\section{Symmetric Solutions}\label{symmetric}

As before, let $\Sigma$ be a smooth hyperelliptic curve of genus $g$, defined by
$$\mu^2=\prod_{j=1}^{2g+2} (\lambda-\lambda_j),$$
where the branch points are symmetric about the real axis, i.e.,
$\lambda_{g+k} = \blamda_k$ for $1\le k \le g+1$.
In this section, we assume that the branch points are also symmetric about the origin.
Then, in addition to the sheet interchange automorphism $\interchange$,
$\Sigma$ has an automorphism
\bel{defofsigma}{\sigma:(\lambda,\mu)\mapsto (-\lambda,(-1)^{g+1}\mu),}
the sign being chosen so that $\sigma$ fixes the points $\infty_+$ and $\infty_-$ on $\Sigma$.
We shall refer to such hyperelliptic curves, and the NLS potentials that arise from them,
as {\it symmetric}.

\subsection{Consequences of Symmetry}
In order to see how the standard homology basis behaves under automorphism $\sigma$,
it is convenient to define an extra cycle $a_{g+1}$, as shown in Figure \ref{genbasis}.
Then, in terms of the $a_j$ for $j\le g$ and the clockwise cycle $\ell$ about $\infty_+$,
\bel{lastacycle}{a_{g+1} \sim -\ell-\sum_{j=1}^g a_j.}
(Here and below, $\sim$ indicates equivalence of basepoint-free homology cycles
within $\Sigma \backslash \{\infty_+,\infty_-\}$.)

\begin{prop}  For $1\le j\le g$, $\sigma(a_j) \sim a_{g+2-j}$, and
$$\sigma(b_j) \sim \left\{\begin{aligned}a_{g+1}+a_1-b_1,\qquad & j=1,\\
b_{g+2-j}+a_1-b_1,\qquad & j>1.\end{aligned}\right.$$
\end{prop}

Let the homology classes be grouped symbolically into row vectors of length $g$:
$$\ula = (a_1,\ldots,a_g), \qquad \ulb = (b_1,\ldots,b_g).$$
Using this notation, and using \eqref{lastacycle} to substitute for $a_{g+1}$,
we may express the results of the proposition in more compact form as
$$\sigma(\ula) \sim \ula M - \ull,\qquad \sigma(\ulb) \sim \ulb\,\Mt + \ula(M-\Mt) - \ull,$$
where $\ull$ is the row vector $(\ell,0,\ldots,0)$, and
$$M=\begin{bmatrix}-1& 0 & 0 & \ldots & 0 & 0 \\ -1 & 0 & 0 & \ldots & 0 & 1 \\
-1 & 0 & 0 & \ldots & 1 & 0 \\ \hdotsfor{6} \\
-1 & 0 & 1 & \ldots & 0 & 0 \\
-1 & 1 & 0 & \ldots & 0 & 0\end{bmatrix}_{g\times g}.$$
(In particular, $M = \begin{bmatrix} -1 & 0 \\ -1 & 1\end{bmatrix}$ when $g=2$.)

\begin{remark}
It is also convenient to group the holomorphic differentials of $\Sigma$ into column vectors.  For example, if the `coordinate differentials'
$\nu_j=(\lambda^{g-j}/\mu)\rd\lambda$, which are holomorphic
for $1\le j \le g$, are grouped into a column vector $\nu$, then the vector of
normalized holomorphic differentials is given by
$$\omega=2\pi \ri \ulA^{-1} \nu,$$
where $\ulA=\nu \cdot \ula$ is the $g\times g$ matrix of period integrals
of the coordinate differentials, and we represent integration of a differential
over a cycle by a dot.
Then the Riemann matrix $\RiemB$ is given by
$$\RiemB = 2\pi \ri \ulA^{-1} (\nu \cdot \ulb).$$
\end{remark}

\bigskip
Using the formula \eqref{defofsigma} and the
transformation of the $a$-cycles under $\sigma$, it is easy to check that the Abelian differentials transform as follows:
$$\sigma^*\rd\Omega_1=-\rd\Omega_1, \quad
\sigma^*\rd\Omega_2=\rd\Omega_2, \quad
\sigma^*\rd\Omega_3=\rd\Omega_3-\omega_1.$$
(For example, $\sigma^*\rd\Omega_3$ has the same asymptotics as $\rd\Omega_3$, but
has period $-2\pi \ri$ on the cycle $a_1$.)  Integrating each of these over the
$b$-cycles, we obtain
\begin{equation}\label{MrVW}
M \vV=-\vV,\quad M \vW=\vW, \quad M \vr=\vr +(\RiemB-2\pi \ri I)_1,
\end{equation}
where the subscript one will denote the first column in a matrix or first entry in a vector.
Similarly, $\sigma^*\omega=M\omega$, giving
\bel{Mtformula}{\RiemB=M \RiemB\, \Mt +2\pi \ri (I-M\Mt),}

Note that $M^2=I$, and $\vV,\vW$ belong respectively to the
$-1$ and $+1$-eigenspaces of $M$, which we will denote as $\M_-$ and $\M_+$.  It is easy to verify that
$\dim \M_- = \lceil g/2 \rceil$ while $\dim \M_+=\lceil (g-1)/2 \rceil$.

Since the symmetric assumption $c=\sum_{j=1}^{2g+2}\lambda_j=0$,
then \eqref{Eformula} from the appendix gives
\begin{align*}
E &=-\dfrac1{\pi \ri}\sum_{j=1}^{g}\int_{a_j} \lambda\omega_j
=-\dfrac1{\pi \ri}\tr\left(\lambda \omega \cdot \ula\right)
=-\dfrac1{\pi \ri}\tr\left(\sigma^*(\lambda\omega) \cdot \sigma(\ula)\right)
\\
&=-\dfrac1{\pi \ri}\tr\left(\sigma^*(\lambda \omega) \cdot (\ula M-\ull)\right)
=\dfrac{\tr\left(\lambda M \omega\cdot \ula M\right)}{\pi \ri} +
    2\tr\left(\sigma^*(\lambda \ulA^{-1}\nu)\cdot \ull\right)\\
&=\dfrac{\tr\left(\lambda M^2 \omega\cdot \ula \right)}{\pi \ri}
 + 2\sum_{j=1}^{g}(\ulA^{-1})_{1j}\int_\ell\sigma^*
 \left( \dfrac{\lambda^{g+1-j}\rd\lambda}{\mu}\right)
=-E+4\pi \ri (\ulA^{-1})_{11}.
\end{align*}
(In the last step, the integral has a residue about $\infty_+$ only when $j=1$.)
Therefore, using \eqref{VWformulas}
\bel{symmetricE}{E=2\pi \ri \left(\ulA^{-1}\right)_{11}=\tfrac12 V_1.}

\subsection{Results}
The following results apply only to filaments reconstructed at the origin, i.e., $\Lambda_0=0$ in the Sym-Pohlmeyer formula.
Note that, with this choice, smooth closure of the filament is not automatic.  When $g$ is even,
the origin is automatically a zero of $\rd\Omega_1$, but it is not automatically a double point.

\begin{theorem}\label{planarthm}
Assume $\Sigma$ is symmetric.  Let $q(s,t)$ be the NLS potential
associated to $(\Sigma,\vD)$, given by \eqref{qformula}.
Let $\Lambda \subset \ZZ^g$ be the one-dimensional lattice of vectors $\vm$ such that $m_1=0$ and
$m_2=\ldots=m_g$ is odd.  (Note that $\Lambda \subset \M_+$.)  Suppose $t_0$ is such that
\bel{specialDcond}{\vD+M \vD-2\ri \vW t_0 = \pi \ri \vm}
for some $\vm\in \Lambda$.  Then $q(s,t_0)$ is the potential for a {\it planar} filament.
\end{theorem}

(Note that, following \S\ref{realityconds}, the vector $\vD$ may be assumed to be purely imaginary.)
\begin{cor}\label{planarcor} If $\Sigma$ is symmetric and $g=2$ or $g=3$,
then $\M_+$ is one-dimensional, and the filament is planar at regular
intervals in time.
\end{cor}

By observing that, in low genus, $\vW$ must be a multiple of $\transpose{\,(0,1,\ldots,1)}\in \M_+$, we deduce that the
gap between planar times is $\pi/W_g$.  It is possible that higher-genus filaments may be planar at isolated times,
but Thm. \ref{planarthm} only implies this if the projection of $\vD$ into $\M_+$ lies along one of a countable
number of lines parallel to $\vW$.  If that is the case, then planarity at repeated times is possible if $\vW$ is a
multiple of $\transpose{\,(0,1,\ldots,1)}$.

\begin{proof}[Proof of Theorem \ref{planarthm}]  Let $\vD'=\vD-\ri \vW t_0$.
To prove planarity at time $t_0$ it suffices to prove that $q_0(s)$, which we define as
\bel{q0eq}{q_0(s):= \dfrac{q(s,t_0)}{A\exp(\ri N t_0)}=
\exp(-\ri E s)\dfrac{\theta(\ri \vV s+\vr-\vD')}{\theta(\ri \vV s-\vD')},
}
is the potential of a planar curve, i.e., the real and imaginary parts of $q_0$, which are the curvatures
for a natural framing along the curve, satisfy a constant-coefficient
homogeneous linear equation (see, e.g., \cite{Bishop}).  To this end, compute
$$\overline{q_0(s)}=\exp(+\ri E s)\dfrac{\theta(-\ri \vV s+\bar{\vr}+\vD')}{\theta(-\ri \vV s+\vD')}
=\exp(\ri E s)\dfrac{\theta(-\ri \vV s+\vr+\vD')}{\theta(\ri \vV s-\vD')},$$
using the fact that $\vD'$ is pure imaginary, $\bar{\vr}\equiv \vr$ mod $2\pi \ri$, and $\theta$ is an even function.
Then, using $\vD'=-M\vD'+\pi \ri \vm$
and \eqref{MrVW}, we have
$$\overline{q_0(s)}=\exp(\ri E s)\dfrac{\theta(M(\ri \vV s+\vr+\RiemB_1-\vD'+\pi \ri \vm))}{\theta(\ri \vV s-\vD')}.$$

The theta-function in the numerator is given by the series
\begin{multline*}
\sum_{\vn\in \ZZ^g}\exp\langle \vn,M(\ri \vV s+\vr+\RiemB_1-\vD'+\pi \ri \vm)+\tfrac12 \RiemB \vn\rangle \\
=\sum_\vn \exp\langle \Mt \vn, \ri \vV s+\vr+\RiemB_1-\vD'+\pi \ri \vm+\tfrac12 M \RiemB \vn\rangle \\
=\sum_\vn \exp\langle \Mt \vn, \ri \vV s+\vr+\RiemB_1-\vD'+\pi \ri \vm +\tfrac12(\RiemB\Mt + 2\pi \ri(M-\Mt)) \vn\rangle \\
=\sum_\vn \exp\langle \Mt \vn, \ri \vV s+\vr+\RiemB_1-\vD'+\tfrac12 \RiemB\Mt \vn\rangle \exp\langle \Mt \vn, \pi \ri (\vm+(M-\Mt) \vn)\rangle,
\end{multline*}
where in the third line we use the substitution $M \RiemB=\RiemB \Mt +2\pi \ri (M-\Mt)$, derived from \eqref{Mtformula}.
Now we observe that the second factor in the last line is equal to one.
For, let $s=n_2+\ldots+n_g$.  Then
$$
\langle \Mt \vn,\vm+(M-\Mt) \vn\rangle  = sm_2+s(-n_1-s)-n_1 s = s(m_2-2n_1-s).
$$
When $s$ is odd, the factor $(m_2-2n_1-s)$ is even, and vice-versa.  Hence, because $\Mt \vn$ ranges over all
points of $\ZZ^g$,
$$\theta(M(\ri \vV s+\vr+\RiemB_1-\vD'+\pi \ri \vm))=\theta(\ri \vV s+\vr+\RiemB_1-\vD').$$

Next, using the quasi-periodicity of $\theta$,
\begin{align*}\overline{q_0(s)}&=\exp(\ri E s)\dfrac{\theta(\ri \vV s+\vr+\RiemB_1-\vD')}{\theta(\ri \vV s-\vD')}
= \exp(\ri E s)\exp(-\ri V_1 s-r_1+D'_1)\dfrac{\theta(\ri \vV s+\vr-\vD')}{\theta(\ri \vV s-\vD')}\\ &=
\exp(-\ri E s)\exp(D'_1-r_1)\dfrac{\theta(\ri \vV s+\vr-\vD')}{\theta(\ri \vV s-\vD')}=\exp(D'_1-r_1)q_0(s),
\end{align*}
using the formula \eqref{symmetricE} for $E$.  Since $q_0(s)$ is a constant multiple of its complex conjugate, it is the
potential of a planar curve.
\end{proof}

\setlength{\tricklength}{.3in}
\begin{figure}[ht]
\centering
\includegraphics[height=.9\textheight, trim = 60 150 60 15]{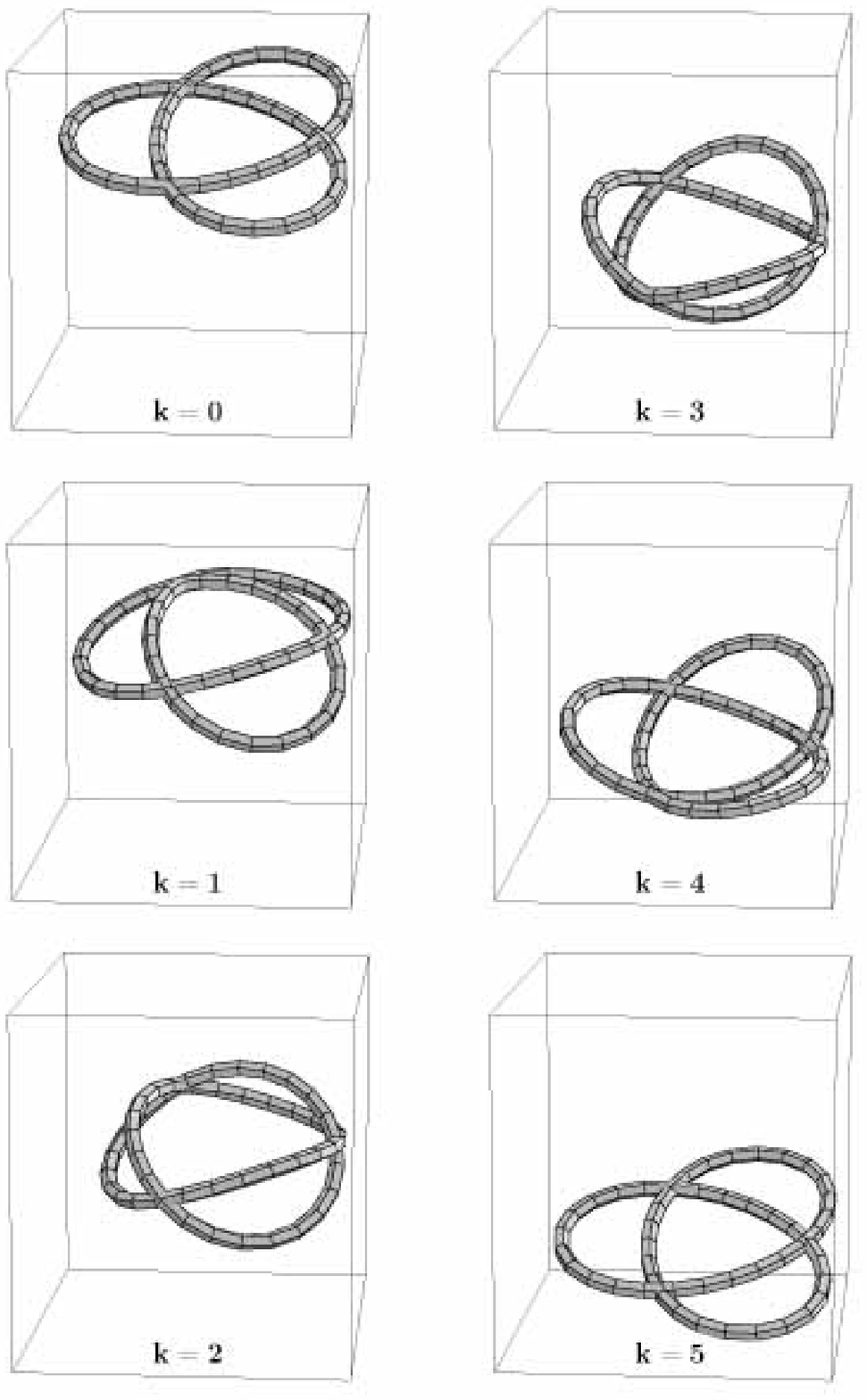}
\caption{Genus 2 symmetric filament, generated using $\vD=0$, at times $t=(.4 k -1)(\pi/2 W_2)$, for $k$ from $0$ to $5$.  This illustrates the periodic planarity of Thm. \ref{planarthm}, the sphericity of Remark \ref{spheremark},
and persistent planes of symmetry of Thm. \ref{twothm}.}\label{pretzel}
\end{figure}

When a filament is planar at regular intervals
(as predicted by Cor. \ref{planarcor}), these planar forms are congruent to
each other.  For, if $t_0$ is a planar
time and we increase time to $t_1=t_0+2\pi/W_g$, then $\vD'$ is decreased by
an integer vector times $2\pi \ri$. The modified potential $q_0$, defined in
\eqref{q0eq}, is unchanged.  So, $q(s,t_1)/q(s,t_0)$ is a unit modulus
constant, and the filament at time $t_2$ is congruent to that at time
$t_0$.  Moreover, the isometry of $\R^3$ that carries the filament at time
$t_0$ to time $t_1$ will be repeated in the evolution from $t_1$ to
subsequent planar times $t_2,t_3$, etc. However, computer experiments in
genus two show that the curve at time $t_{1/2}=t_0 + \pi/W_g$ is also
congruent to the curve at time $t_0$, but this doesn't happen in genus
three.

\begin{remark}\label{spheremark}
Another phenomenon (which so far is only experimentally verified) is that
in genus two, between the planar times $t_0, t_{1/2}, t_1,\ldots$, the
curve lies on a sphere whose center and radius vary with time.  (This can
be checked numerically by evaluating the real and imaginary parts of the
potential $q(s,t)$ at a fixed time, and verifying that they satisfy an
inhomogeneous linear equation.)  Vortex filament flow solutions obtained
as B\"cklund  transformations of translating circles also lie on spheres
\cite{Ca1}.
\end{remark}

\def\gt{\widetilde{\vgamma}}
\begin{prop}\label{symprop} Let $q(s)$ be the potential for a curve $\vgamma(s)$ in $\R^3$.
If $q(-s)=q(s)$, then the curve is symmetric under reflection in the normal plane through $\vgamma(0)$.
If $\overline{q(s)}=q(-s)$, then the curve is symmetric under
a 180-degree rotation about the normal line through $\vgamma(0)$.
\end{prop}
\begin{proof}
Let $T(s),N_1(s),N_2(s)$ be an oriented natural frame along $\vgamma$, with natural curvatures
$k_1,k_2$ which are the real and imaginary parts of $q(s)$ respectively.
Let $\gt(s)=\vgamma(-s)$.
Then in the first case $(-T(-s),N_1(-s),N_2(-s))$ is an
oppositely-oriented natural frame for $\gt$ with the same curvature
functions as the given oriented frame for $\vgamma$.  Hence $\gt(s)$ must
be congruent to $\vgamma(s)$ under an orientation-reversing isometry.
Since that motion must fix $\vgamma(0)$ and $N_1(0),N_2(0)$, then it must
be reflection in the normal plane through $\vgamma(0)$. In the second case,
$(-T(-s),N_1(-s),-N_2(-s))$ is a similarly-oriented natural frame for
$\gt$ with the same curvatures, and $\gt(s)$ must be congruent to
$\vgamma(s)$ under an orientation-preserving isometry that fixes
$\vgamma(0)$ and the line through $\vgamma(0)$ along $N_1(0)$.  (Since
$k_2(0)=0$, this is the normal line at $\vgamma(0)$.)
\end{proof}

\begin{prop} Let $\Sigma$ be symmetric, and assume there are $s_0$ and $t_0$ such that
\begin{equation}\label{rotDcond}
\vD - (\ri \vV s_0 + \ri \vW t_0) = \pi \ri \vk
\end{equation}
for some vector $\vk$ of odd integers.
Then the filament is symmetric under
a 180-degree rotation about the normal line through $\vgamma(s_0,t_0)$
\end{prop}
\begin{proof}
Let $s'=s-s_0$, and let
$$q_0(s')=\dfrac{q(s,t_0)}{A \exp(-\ri E s_0  + \ri N t_0)}
=\exp(-\ri E s')\dfrac{\theta(\ri \vV s'+\vr-\pi \ri \vk)}{\theta(\ri \vV s'-\pi \ri \vk)}$$
It is easy to see that
$$\widebar{q_0(s')}=\exp(+\ri E s')\dfrac{\theta(-\ri \vV s'+\bar{\vr}+\pi \ri \vk)}{\theta(-\ri \vV s'+\pi \ri \vk)}
=\exp(\ri E s')\dfrac{\theta(-\ri \vV s'+\vr-\pi \ri \vk)}{\theta(-\ri \vV s'-\pi \ri \vk)}=q_0(-s'),$$
using the $2\pi \ri$-periodicity of $\theta$.
Since $|A|q_0(s')$ is a potential for the filament $\vgamma(s,t_0)$,
 then the result follows by Prop. \ref{symprop}.
\end{proof}

\begin{theorem}\label{twothm}
Assume $\Sigma$ is symmetric and of genus 2.  Then there is a fixed plane in $\R^3$ such that
at each time the filament is symmetric under reflection in that plane.
At each `planar time' $t_0$ given
by Thm. \ref{planarthm}, the plane containing the filament is perpendicular to the
plane fixed by the reflection.
\end{theorem}

\begin{proof}
Let $t_0$ be a planar time, i.e., suppose \eqref{specialDcond}
is true for some $\vm\in \Lambda$.  Then there exist $s_0$ and an odd integer
vector $\vk$ such that \eqref{rotDcond} is satisfied.
For, \eqref{specialDcond} says that the projection of $\vD'=\vD-\ri \vW t_0$ into $\M_+$ is
$\tfrac12 \vm$.  Given this, then \eqref{rotDcond} implies that
$\vk+M \vk=\vm$, which amounts to
$2k_2 - k_1 = m_2$.  Given any $\vk$ that satisfies this condition, there
exists an $s_0$ that satisfies \eqref{rotDcond}, because $\vV$ spans $\M_-$.

Thus, the curve is
symmetric under rotation $\calT$ about an axis $L$ through $\vgamma(s_0,t_0)$.
However, since the filament is also planar at time $t_0$, the rotation $\calT$ can
be replaced by the reflection $\calR$ in the plane through $L$ perpendicular to the plane
containing the filament.

To complete the proof, we use the symmetry properties of the vortex
filament flow $$\vgamma_t=\vgamma_s \times \vgamma_{ss}.$$ If $\vgamma(s,t)$
is a solution, and $\calT$ is a orientation-preserving isometry of $\R^3$,
then it is easy to check that $\gt(s,t):=\calT(\vgamma(s,t))$ is
another solution.  In particular, if $\gt(s,t_0)=\vgamma(s-c,t_0)$
for some $c$ and some time $t_0$, then this equation holds at all times.
In other words, symmetry under $\calT$ is preserved by the flow.

Similarly, if $\calR$ is an orientation-reversing isometry, then
$\gt(s,t):=\calR(\vgamma(-s,t))$ is another solution.  Since
$\vgamma$ is symmetric under reflection $\calR$ at time $t_0$, and this
symmetry reverses orientation along $\vgamma$, then
$\calR(\vgamma(-s,t))=\vgamma(s-c,t_0)$ for some $c$, and it follows that
this relationship persists in time.
\end{proof}

\setlength{\tricklength}{.4in}
\newlength{\scrolength}
\setlength{\scrolength}{.2in}
\begin{figure}[ht]
\centering
\includegraphics[height=.9\textheight, trim = 60 150 60 15]{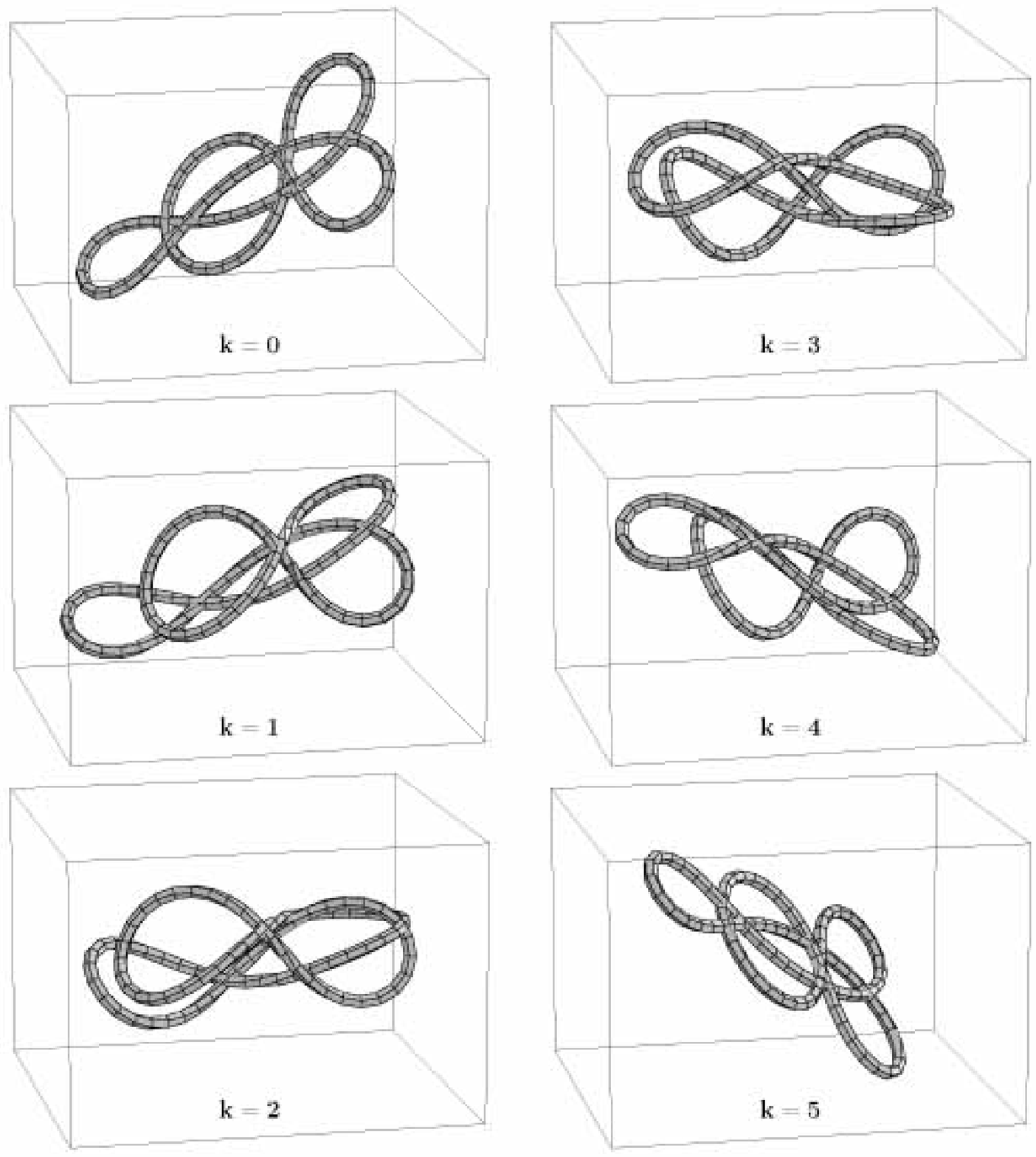}
\caption{Genus 3 symmetric filament, generated using $\vD=0$,
at times $t=(.4 k -1)(\pi/2 W_2)$, for $k$ from $0$ to $5$..  Note the persistent self-intersections in the plane of reflection symmetry.
The curve returns to its original shape, in a rotated plane.}\label{scissors}
\end{figure}

\begin{remark} At a planar time,
a filament must cross the axis of symmetry at least once, since
the filament is connected to its reflection.  If it crosses the axis in a direction
that is not perpendicular to the axis, a self-intersection occurs at this point.
The fact that reflection symmetry persists as the curve lifts out of the plane
implies that these self-intersections also persist (see Figures \ref{pretzel} and
\ref{scissors}).  However, it is possible that, as the
curve evolves, two self-intersections may coalesce and disappear, while maintaining reflection
symmetry.

Note that not all planar curves that are symmetric about an axis have self-intersections
which are preserved by the flow.  It is easy to draw examples where the reflection
symmetry preserves the orientation along the curve (rather than
reversing it, which is the property used in the proof of Theorem \ref{twothm}).
Then, under the filament flow, the strands immediately pull away in
opposite directions from the self-intersections.  In this case, the
symmetry is the restriction of an orientation-preserving 180-degree
rotation in $\R^3$, and such rotational symmetry persists as the filament
evolves.
\end{remark}

\begin{remark}  The argument in the proof of Theorem
2 relies on the fact that, when $g=2$, $\vV$ and $\vW$ span $\R^g$.
However, given a symmetric hyperelliptic curve of genus $g\ge 3$, there
are plenty of potentials to which the argument in the proof of Theorem \ref{twothm} applies (see Figure \ref{scissors}).
For, given any vector $\vk$ of odd integers, we may choose an
imaginary vector $\vD$ so that \eqref{rotDcond} is satisfied for some $s_0$ and $t_0$.
Then $\vD$ also satisfies \eqref{specialDcond} if and only if
the differences $k_{g-j} - k_{2+j}$, $j\ge 0$, are all equal.
The latter condition is vacuous when $g=3$.
Thus, in genus 3 the set of appropriate divisors corresponds to the
image of the affine lattice of odd integer vectors in the quotient vector
space $\R^3 / \{\vV,\vW\}$.
(Changing $\vD$ by a multiple of $\vV$ or $\vW$ merely
corresponds to a shift in $s_0$ or $t_0$.)
\end{remark}
\clearpage

`\section{Appendix}\label{appendix}
\def\cL{\mathcal{L}}

In this appendix, we will indicate how the formulas for finite-gap NLS Baker-Akhiezer
functions developed in \S\ref{thetasection} are related to those given in the
monograph by Belokolos, Bobenko, Enol'skii, Its and Matveev \cite{BBEIM}.  We will also
develop (from a slightly different perspective than \cite{BBEIM})
the reality condition which must be satisfied by the divisor,
and discuss how to compute some of the quantities
used in the expression for the finite-gap NLS potential.

\subsection{Alternative Eigenfunction Formulas}\label{reconciliation}
From \S\ref{eigenfunctionsection}, we had
\begin{multline*}
\psi^1(P;s,t) = \dfrac{g_+(P)}{g_+(\infty_-)} \exp\left(\ri s(\Omega_1(P)-\tfrac E2) + \ri t(\Omega_2(P)+\tfrac N2)\right)\\
\times \dfrac{\theta(\A(P) + \ri \vV s +\ri \vW t -\A(\cD_+)-\K)\,\theta(\A(\infty_-)-\A(\cD_+)-\K)}
{\theta(\A(\infty_-) + \ri \vV s +\ri \vW t -\A(\cD_+)-\K)\,\theta(\A(P)-\A(\cD_+)-\K)}
\end{multline*}
and
\begin{multline*}
\psi^2(P;s,t) = \dfrac{g_-(P)}{g_-(\infty_+)} \exp\left(\ri s(\Omega_1(P)+\tfrac E2) + \ri t(\Omega_2(P)-\tfrac N2)\right)\\
\times \dfrac{\theta(\A(P) + \ri \vV s +\ri \vW t -\A(\cD_-)-\K)\,\theta(\A(\infty_+)-\A(\cD_-)-\K)}
{\theta(\A(\infty_+) + \ri \vV s +\ri \vW t -\A(\cD_-)-\K)\,\theta(\A(P)-\A(\cD_-)-\K)}.
\end{multline*}

Recall that $\rd\Omega_3$ is the Abelian differential on $\Sigma$, with zero $a$-periods, such that
$$\rd\Omega_3 \sim \pm \dfrac{\rd\lambda}{\lambda}\qquad \text{as}\ \lambda \to \infty_\pm.$$
Then $\oint_\ell \rd\Omega_3 = 2\pi \ri$, where $\ell$ is the clockwise cycle around $\infty_+$ shown in
Figure \ref{genbasis}.
\begin{lemma}\label{rrelationship}
$\oint_{b_j} \rd\Omega_3 = -r_j$, where $r_j$ denotes the $j$th entry of vector $\vr$.
\end{lemma}
This will be proved at the end of \S\ref{practicalc}.  For now,
we note that the corresponding Abelian integral $\Omega_3$
behaves like $\pm \log(\lambda)$ near $\infty_\pm$.
Furthermore, we observe that the quotient
$$f(P)=\exp(\Omega_3(P)) \dfrac{\theta(\A(P)-\vD-\vr)}{\theta(\A(P)-\vD)}$$
is a well-defined meromorphic function on $\Sigma$.  For, if the paths of integration
in the integrals $\Omega_3(P)$ and $\A(P)$
are changed by the addition of the same $a$-cycle, each factor in $f(P)$ is unchanged;
if the cycle $b_j$ is added, then $f(P)$ is modified by the factor
$$\exp\left(\oint_{b_j}\rd\Omega_3\right)\exp(r_j)=1.$$

Note that if, say, we confine the path of integration for $\Omega_3$ to the interior of
the cut Riemann surface $\Sigma_0$ described in \S\ref{eigenfunctionsection},
then $\exp(\Omega_3(P))$ has a pole at $\infty_+$ and a zero at $\infty_-$.
Consequently, $f(P)$ has pole divisor $\cD_+ + \infty_+$ and zero divisor $\cD_- + \infty_-$.
It follows that
\bel{constexp}{\exp(\Omega_3(P)) \dfrac{\theta(\A(P)-\vD-\vr)}{\theta(\A(P)-\vD)}\times \dfrac{g_+(P)}{g_-(P)}}
is a nonzero constant.

We can express this constant in terms of the asymptotic
behavior of $\Omega_3$ and $g_+$.  Namely, suppose $P$ is near $\infty_+$ and lies over
the real axis in the $\lambda$-plane.  Let $C_P$ be a path from $P$ to $\interchange(P)$
in the interior of $\Sigma_0$, such that $C_P$ is invariant under sheet interchange.
Let $a_{g+1}$ be the cycle around the rightmost branch cut, which satisfies the
homological equivalence \eqref{lastacycle}
within $\Sigma - \{\infty_+, \infty_-\}$.  Then we have
$$\int_{C_P} \overline{\rd\Omega_3} = \int_{\tau(C_P)} \rd\Omega_3 =
\int_{C_P+a_{g+1}}\rd\Omega_3 = -2\pi \ri + \int_{C_P} \rd\Omega_3,$$
where $\tau:(\lambda,\mu) \mapsto (\blamda,\overline{\mu})$ is complex conjugation on $\Sigma$.
Therefore,
\bel{betalimit}{
\lim_{P\to \infty_+} \left(\int_{C_P}\rd\Omega_3 - 2 \log \lambda(P)\right) = \pi \ri - \log\beta
}
for some positive real number $\beta$.  (In terms of the notation in \cite{BBEIM}, the
above limit is written as $-\log \omega_0$ for a negative real number $\omega_0$.)
Therefore,
$$\exp(\Omega_3(P)) \sim \dfrac{\ri}{\sqrt{\beta}} \lambda \qquad \text{as}\ P \to \infty_+.$$
Suppose that $g_+(P) = \dfrac{1}{\alpha \lambda} + O(\lambda^{-2})$ when $P$ is near $\infty_+$,
 for some constant $\alpha$.
Then letting $P$ tend to $\infty_+$ in \eqref{constexp} gives
$$\exp(\Omega_3(P)) \dfrac{\theta(\A(P)-\vD-\vr)}{\theta(\A(P)-\vD)}\times \dfrac{g_+(P)}{g_-(P)}
=\dfrac{\ri}{\alpha \sqrt{\beta}} \dfrac{\theta(\vD)}{\theta(\vD-\vr)}.$$
Solving this equation for $g_-(P)$ and substituting in $\psi^2$ gives
\begin{multline*}
\psi^2 = g_+(P) \dfrac{\alpha \sqrt{\beta}}{\ri} \exp\left(\ri s(\Omega_1(P)+\tfrac E2) + \ri t(\Omega_2(P)-\tfrac N2)+\Omega_3(P)\right)\\
\times  \dfrac{\theta(\A(P) + \ri \vV s +\ri \vW t -\vD-\vr)\,\theta(\vD-\vr)}{\theta(\ri \vV s +\ri \vW t -\vD)\,\theta(\A(P)-\vD)},
\end{multline*}
which agrees with the formula 4.1.16 for $\psi^2$ in \cite{BBEIM}, up to the multiplicative
factor of $g_+(P)$.  Similarly, $\psi^1$ coincides with the formula for $\psi^1$ in \cite{BBEIM} except for
multiplication by $g_+(P)$.  Since our vector-valued function and that of
\cite{BBEIM} differ only by multiplication
by a scalar that is independent of $s$ and $t$, both satisfy the NLS linear system for the same potential $q(s,t)$.

\subsection{Reality Conditions}\label{realityconds}

By inserting the matrix $\Psi$ defined by \eqref{psipsibarmatrix} into the NLS linear system,
and expanding near $\infty_-$ in powers of $\lambda$, we
obtain
\bel{qformula}
{q(s,t) = A \exp(-\ri E s +\ri N t)\dfrac{\theta(\ri \vV s+\ri \vW t-\vD+\vr)}{\theta(\ri \vV s + \ri \vW t -\vD)}
\qquad \text{where}\
A = \dfrac{2\theta(\vD)}{\alpha \theta(\vD-\vr)},}
and
\bel{qbarformula}
{\overline{q(s,t)} = \dfrac{4\beta}{A} \exp(\ri E s - \ri N t)\dfrac{\theta(\ri \vV s + \ri \vW t - \vD-\vr)}{\theta(\ri \vV s + \ri \vW t-\vD)}.}

It is not immediate that the second expression is the complex conjugate of the first; in fact, the
reality conditions on $\vD$ come from imposing this condition.
Before computing this, we need to calculate the imaginary parts of the members of \eqref{qformula}.
Since $\tau^* \rd\Omega_j = \overline{\rd\Omega_j}$ and
$$\tau(b_k) \sim b_k + \ell +\sum_{l \ne k} a_l$$
within $\Sigma -\{\infty_+, \infty_-\}$,
then $\vV$ and $\vW$, being the $b$-periods of $\rd\Omega_1$ and $\rd\Omega_2$, are real vectors.
Similarly, Lemma \ref{rrelationship} implies that
$$r_k = \overline{r_k} + 2\pi \ri.$$
Moreover, since $\tau^*\omega_k = \overline{\omega_k}$, then
the Riemann matrix $\RiemB$, defined by \eqref{pmatrix} satisfies
\begin{equation}\label{Bform}
\overline{\RiemB_{kl}} = \left\{ \begin{aligned} \RiemB_{kl} &\text{ for}\ k=l \\
\RiemB_{kl} + 2\pi \ri &\text{ for}\ k\ne l.\end{aligned} \right.
\end{equation}
It follows that $\overline{\theta(z)} = \theta(\overline{z}).$

Now comparing \eqref{qformula} and \eqref{qbarformula} gives the reality condition
$$\overline{A} \dfrac{\theta(-\ri \vV s - \ri \vW t - \overline{\vD}+ \overline{\vr})}{\theta(-\ri \vV s-\ri \vW t - \overline{\vD}}
= \dfrac{4\beta}{A} \dfrac{\theta(\ri \vV s + \ri \vW t - \vD-\vr)}{\theta(\ri \vV s + \ri \vW t-\vD)}.$$
Since poles and zeros must match up, we must have
\bel{Dcond}{
\vD+\overline{\vD} = 2\pi \ri {\vn} + \RiemB {\vm},}
where $\vm,\vn \in \ZZ^g$ are such that
\bel{Acond}{|A|^2 \exp(\langle \vm,\vr\rangle) = 4\beta.}
Taking the imaginary part of \eqref{Dcond}, with \eqref{Bform} taken into account, gives
\bel{realitymn}{2n_j=\sum_{k\ne j} m_k.}
(In particular, $\vn=0$ when $g=1$.)
Thus, all the entries of $\vm$ have the same parity, even or odd.
When $g$ is even, summing \eqref{realitymn} over $j$ shows that $\vm$ must be all even.
If $g$ is odd and $\vm$ is all odd, the fact that $\Im(r_j)=\pi$ would imply that the left-hand side of
\eqref{Acond} is negative.  Thus, we can assume that $\vm$ is all even.
In this case, one can let $\vD'=\vD-\tfrac12 \RiemB \vm$; then $\vD'$ is purely imaginary,
and \eqref{qformula},\eqref{Acond} are still valid with $\vD$ replaced by $\vD'$ and $\vm$ replaced by zero.  We now do this, and
then our reality condition amounts to the vector $\vD$ being purely imaginary:
\bel{simpleDcond}{\overline{\vD} = -\vD.}

We can, without loss of generality, assume that $A$ is real and positive; then \eqref{Acond}
gives $A = 2\sqrt{\beta}$.  From \eqref{qformula} we then get
$\alpha\sqrt{\beta}=2\theta(\vD)/\theta(\vD-\vr)$.  So, omitting the common
scalar multiples $g_+(P)\theta(\vD)/\theta(\A(P)-\vD)$ from the eigenfunctions gives the
{\em simplified formulas}
\begin{align*}
\psi^1 &= \exp\left(\ri s(\Omega_1(P)-\tfrac E2) + \ri t(\Omega_2(P)+\tfrac N2)\right)
 \dfrac{\theta(\A(P) + \ri \vV s +\ri \vW t -\vD)}
{\theta(\ri \vV s +\ri \vW t -\vD)}, 
\\
\psi^2 &= -\ri\exp\left(\ri s(\Omega_1(P)+\tfrac E2) + \ri t(\Omega_2(P)-\tfrac N2)\right)
\dfrac{\theta(\A(P) + \ri \vV s +\ri \vW t -\vD-\vr)}{\theta(\ri \vV s +\ri \vW t -\vD)}.
\end{align*}

\bigskip
We will now interpret the reality condition \eqref{simpleDcond}
in terms of the divisors $\cD, \cD_+, \cD_-$ on $\Sigma$.

\def\itau{\interchange\tau}
\def\Dzero{\cD}
\begin{prop}If $\overline{\vD} = -\vD$, then $\cD_- = \itau(\cD_+)$.
\end{prop}
\noindent
(Recall that $\interchange$ denotes the sheet interchange automorphism
and $\tau$ denotes complex conjugation on $\Sigma$.)

\begin{proof} The function $f(P) = \theta(\A(P) - \vD)$ has zero divisor $\cD_+$.  (To make $f$
well-defined, we restrict to the cut Riemann surface $\Sigma_0$, assuming that the homology basis avoids the points of $\cD_+$ and $\cD_-$.)  Therefore, the function
$$\overline{f(\interchange\tau(P))} = \theta\left(\overline{\A(\itau(P))} + \vD\right)$$
has zero divisor $\interchange\tau(\cD_+)$.
But since $\infty_+ = \itau(\infty_-)$ and
$(\itau)^*\omega_j = -\overline{\omega_j}$, then
$$
\A(\itau(P)) = \int_{\infty_-}^{\infty_+} \omega + \int_{\itau(\infty_-)}^{\itau(P)} \omega \\
= \vr - \overline{\A(P)}.$$
Using the fact that $\overline{r_j} \equiv r_j$ modulo $2\pi \ri$,
$$\theta\left(\overline{\A(\itau(P))} + \vD\right) = \theta(\overline{\vr} - \A(P) + \vD) =
\theta(\A(P)-\vD-\vr) = \theta(\A(P) - \A(\cD_-) - \K),$$
which has $\cD_-$ as its zero divisor.
\end{proof}

\begin{prop}\label{realiff}
Assuming $\Dzero$ satisfies the Genericity Condition of \S\ref{thetasection}, then
$\cD_- = \itau(\cD_+)$ if and only if $\itau(\Dzero)\sim \Dzero$, where $\sim$
denotes linearly equivalence of divisors.
\end{prop}

Before proving the lemma, we recall some pertinent facts about divisors on a Riemann surface (see, e.g., \cite{Du} or \cite{kirwan}).
For any divisor $\cD$, the meromorphic
functions $f$ such that $\cD + (f)\ge 0$ form a vector space $\cL(\cD)$ with dimension
$l(\cD)$.  For example, a positive divisor always has $l(\cD)\ge 1$, because of the constant
functions.  If two divisors $\cD,\cD'$ are linearly equivalent, then $l(\cD)=l(\cD')$.

The Genericity Condition can be stated as
\bel{hzeroassumption}{l(\Dzero-\infty_- -\infty_+)=0.}
This implies that $l(\cD_+)=l(\Dzero - \infty_+) = 1$, and similarly for $\cD_-$
and $\Dzero - \infty_-$.  For, if $l(\Dzero -\infty_+)>1$, let
$f_1,f_2 \in \cL(\Dzero-\infty_+)$ be linearly independent; then there exists a nontrivial
linear combination $c_1 f_1 + c_2 f_2$ that vanishes at $\infty_-$, and this
contradicts \eqref{hzeroassumption}.

\begin{proof}[Proof of Prop. \ref{realiff}]
Suppose that $\cD_- = \itau(\cD_+)$.  By construction,
$$\cD_+ = \Dzero - \infty_+ + (f)$$
for some meromorphic function $f$.  Appying $\itau$ to both sides gives
$$\cD_- = \itau(\Dzero) - \infty_- + \left(\overline{f \circ \itau}\right).$$
Since $\cD_- \sim \Dzero -\infty_-$, then $\Dzero\sim \itau(\Dzero)$.

The same computation, run backwards, shows that if $\Dzero \sim \itau (\Dzero)$ then
$\cD_- \sim \itau(\cD_+)$, i.e.,
$$\cD_- + (f) = \itau(\cD_+)$$
for some meromorphic function $f$.
However, $l(\cD_-)=1$ implies that $\cD_- +(f) \ge 0$ only when $f$ is a constant.  Thus,
$\cD_- = \itau(\cD_+)$.
\end{proof}

\subsection{Computing $E$, $N$, $\vr$ and $\beta$}\label{practicalc}
Suppose functions $\Omega,\Omega'$ are holomorphic on the cut Riemann surface $\Sigma_0$.
Then by Stokes' Theorem,
\bel{mrstokes}{0 = \int_{\Sigma_0} \rd(\Omega' \rd\Omega) = \oint_{\partial \Sigma_0} \Omega' \rd\Omega =
\sum_{j=1}^g A'_j B_j - B'_j A_j,}
where $A_j = \oint_{a_j} \Omega$, $B_j = \oint_{b_j} \Omega$, and similarly for the primes.
(See, for example, \cite{Du}.)
We will use this relationship to derive formulas for $E$ and $N$
and to prove Lemma \ref{rrelationship}.  (Our formula for $E$ is equivalent
to one given without proof in \cite{BBEIM}.)
The differentials we use will actually have singularities at $\infty_{\pm}$, so the left-hand
side of \eqref{mrstokes} will be modified by adding the integrals
of $\Omega' \rd\Omega$ along clockwise cycles about these points:
\bel{morestokes}{
0=\sum_{j=1}^g A'_j B_j - B'_j A_j+\oint_{\ell + \interchange(\ell)} \Omega' \rd\Omega.}
Each of these extra terms
will be $-2\pi \ri$ times the residue of the differential at the singularity.

First, let $\Omega'=\Omega_1$ and $\rd\Omega = (\lambda^g /\mu) \rd\lambda$.  By expanding
$\mu^2$ in terms of $\lambda$ and applying the binomial series, we get that, near $\infty_{\pm}$,
$$\dfrac{\mu}{\lambda^{g+1}} =
\pm\left(1 - \frac{c}2\lambda^{-1} - d \lambda^{-2}+ O(\lambda^{-3})\right),$$
where $c=\sum_{j=1}^{2g+2} \lambda_j$ and
\bel{dformula}{
d=-\frac18\left(c^2-2\sum_{j=1}^{2g+2} \lambda_j^2 \right)
}
This, together with
$$\Omega_1 = \pm (\lambda + \frac{E}2 + o(1))$$
allows us to expand the integrand near $\infty_\pm$:
$$\Omega' \rd\Omega =
-\lambda^2\left( 1 - \left(\dfrac{E-c}2\right)\lambda^{-1} + O(\lambda^{-2})\right)
\rd(\lambda^{-1}).$$
Since $\rd\Omega'=\rd\Omega_1$ has no $a$-periods, from \eqref{morestokes} we have
$$ 0 = -2\pi \ri(E-c) - \sum_{j=1}^g V_j \oint_{a_j} \dfrac{\lambda^g}\mu \rd\lambda.$$
Thus,
\bel{Eformula}{
E = c - \dfrac1{2\pi \ri} \sum_{j=1}^g V_j \oint_{a_j} \dfrac{\lambda^g}\mu \rd\lambda.
}

Next, let $\Omega'=\Omega_2$ instead.  Then
$$\Omega' \rd\Omega =
-\lambda^3\left( 2 + c\lambda^{-1}+ \left(\dfrac{N}2+2d + \dfrac{c^2}2\right)+ O(\lambda^{-3})\right)
\rd(\lambda^{-1})$$
near $\infty_{\pm}$, and \eqref{morestokes} gives
$$0 = -2\pi \ri \left(N+4d + c^2\right)-\sum_{j=1}^g W_j \oint_{a_j} \dfrac{\lambda^g}\mu \rd\lambda,$$
so that
\bel{Nformula}
{N = -4d - c^2 - \dfrac{1}{2\pi \ri}\sum_{j=1}^g W_j \oint_{a_j} \dfrac{\lambda^g}\mu \rd\lambda.}

We note that, by letting $\rd\Omega = \omega_j$ and $\Omega' = \Omega_1$ or $\Omega_2$, similar
calculations give
\bel{VWformulas}{0 = 2c_{j1} - V_j, \qquad 0 = 4c_{j2} + 2c\, c_{j1} - W_j,}
where $c_{jk}$ are the constants such that 
$$\omega_j = \sum_{k=1}^g c_{jk} \dfrac{\lambda^{g-k}}\mu \rd\lambda.$$
Thus, computing just the $a$-periods of the coordinate differentials
$\nu_k =\dfrac{\lambda^{g-k}}\mu \rd\lambda$ for $0\le k \le g$,
we can determine $E,N$ and
the $b$-periods of $\rd\Omega_1,\rd\Omega_2$ algebraically.

In most experiments we have carried out, it is straightforward to calculate
the $a$-periods of the coordinate differentials by numerical integration.
Numerical techniques are also practical for calculating $\beta$
 in the limit \eqref{betalimit}.  For this purpose, we can rewrite that
 limit as
$$2\pi \ri - \log \beta = 2 \log(\lambda_{2g+2}) + 2 \lim_{P \to \infty_+}
\int_{\Gamma_P} \left(\rd\Omega_3 - \lambda^{-1}\rd\lambda\right),$$
where $\Gamma_P$ is the portion of $C_P$ from $\lambda_{2g+2}$ to $P$
on the upper sheet.  Now the improper integral converges.

\begin{proof}[Proof of Lemma \ref{rrelationship}]
If we let $\Omega'=\Omega_3$ and $\rd\Omega=\omega_j$,
we must make additional cuts.
Let $\Sigma'_0$ be the cut Riemann surface $\Sigma_0$ (shown in Figure \ref{cutting Sigma}) with an additional cut from $\infty_-$
to $\infty_+$, along $\Pi \cup \interchange(\Pi)$.  Then $\Omega_3$ is well-defined
and holomorphic on $\Sigma'_0$,
with its values of $\Omega_3$ differing by $2\pi \ri$ on each
side of this cut.  From \eqref{mrstokes} we obtain
$$0 = -2\pi \ri \oint_{b_j} \rd\Omega_3 - 2\pi \ri \int_{\infty_-}^{\infty_+} \omega_j.$$
\end{proof}

\newcommand{\ditto}{{\leavevmode\vrule height 2pt depth -1.6pt width 23pt\,}}

\end{document}